\documentclass[reprint, prl, aps, floatfix, nofootinbib, superscriptaddress]{revtex4-2}
\usepackage{amsmath}
\usepackage{rotating}
\usepackage{graphicx}
\usepackage{chemformula}
\usepackage{mathtools}
\usepackage{tabularx}
\usepackage{color}
\usepackage{bm}
\usepackage{dcolumn}
\usepackage[colorlinks=true, linkcolor=blue, citecolor=blue, urlcolor=blue]{hyperref}
\usepackage{soul}

\begin{document}

\title{Magnetic ground state of a prototype quasicrystal approximant: a candidate for octahedral spin ice physics}
\author{Leonie Woodland}
\thanks{These authors contributed equally to this work.}
\affiliation{ISIS Facility, Rutherford Appleton Laboratory, Chilton, Didcot OX11 0QX, UK}
\affiliation{Clarendon Laboratory, University of Oxford Physics Department, Parks Road, Oxford OX1 3PU, UK}
\author{Farid Labib}
\thanks{These authors contributed equally to this work.}
\affiliation{Research Institute for Science and Technology, Tokyo University of Science, Tokyo 125-8585, Japan}
\author{Dmitry Khalyavin}
\affiliation{ISIS Facility, Rutherford Appleton Laboratory, Chilton, Didcot OX11 0QX, UK}
\author{Pascal Manuel}
\affiliation{ISIS Facility, Rutherford Appleton Laboratory, Chilton, Didcot OX11 0QX, UK}
\author{Fabio Orlandi}
\affiliation{ISIS Facility, Rutherford Appleton Laboratory, Chilton, Didcot OX11 0QX, UK}
\author{Hubertus Luetkens}
\affiliation{Laboratory for Muon-Spin Spectroscopy, Paul Scherrer Institut, CH-5232 Villigen PSI, Switzerland}
\author{Ryuji Tamura}
\affiliation{Department of Materials Science and Technology, Tokyo University of Science, Tokyo 125-8585, Japan}
\date{\today}

\begin{abstract}
    Magnetic ordering in quasicrystals has recently emerged as a fertile ground for discovering unconventional magnetic states beyond the framework of periodic crystals. However, elucidating the microscopic origin of such states remains challenging due to the intrinsic aperiodicity of quasicrystals. Here, we address this issue by investigating the prototypical Tsai-type quasicrystal approximant Cd$_6$Tb, which preserves the essential local geometry and connectivity of icosahedral quasicrystals while allowing detailed structural and magnetic characterization due to its translational periodicity. Using neutron diffraction measurements, we find a noncoplanar multi-\textit{k} magnetic ground state composed of Ising-like Tb moments arranged on a network of corner-sharing octahedra, the ingredients required to host octahedral spin-ice physics. Remarkably, only one third of the Tb moments develop long-range magnetic order, whereas the remaining moments display strongly reduced static order accompanied by persistent spin dynamics on microsecond timescales, as evidenced by muon spin rotation. This coexistence of ordered and fluctuating moments constitutes a potential realization of magnetic fragmentation --- a key prediction of octahedral spin-ice physics --- in a quasicrystal-related material.

\end{abstract}

\maketitle

Quasicrystals, characterized by long-range order without translational periodicity \cite{Shechtman1984,Shechtman1985}, challenge conventional paradigms of condensed-matter physics and continue to reveal unexpected collective phenomena.
While most magnetic quasicrystals form spin glasses \cite{Fisher1999,Sato2001,Kong2014,AlQadi2009}, as might be expected from the frustration inherent to the icosahedral geometry, long-range ferromagnetic \cite{Tamura2021,Takeuchi2023} and antiferromagnetic \cite{Tamura2025} orders have recently been experimentally observed many years after being shown to be theoretically possible \cite{Lifshitz1998}. Alongside recent theoretical predictions of emergent magnetic states inaccessible in periodic crystals \cite{Thiem2015a,qubit,Watanabe2021},
these developments have transformed quasicrystals from structural curiosities into conceptually important systems for exploring new forms of correlated matter.

\begin{figure}
    \centering
    \includegraphics[width=\linewidth]{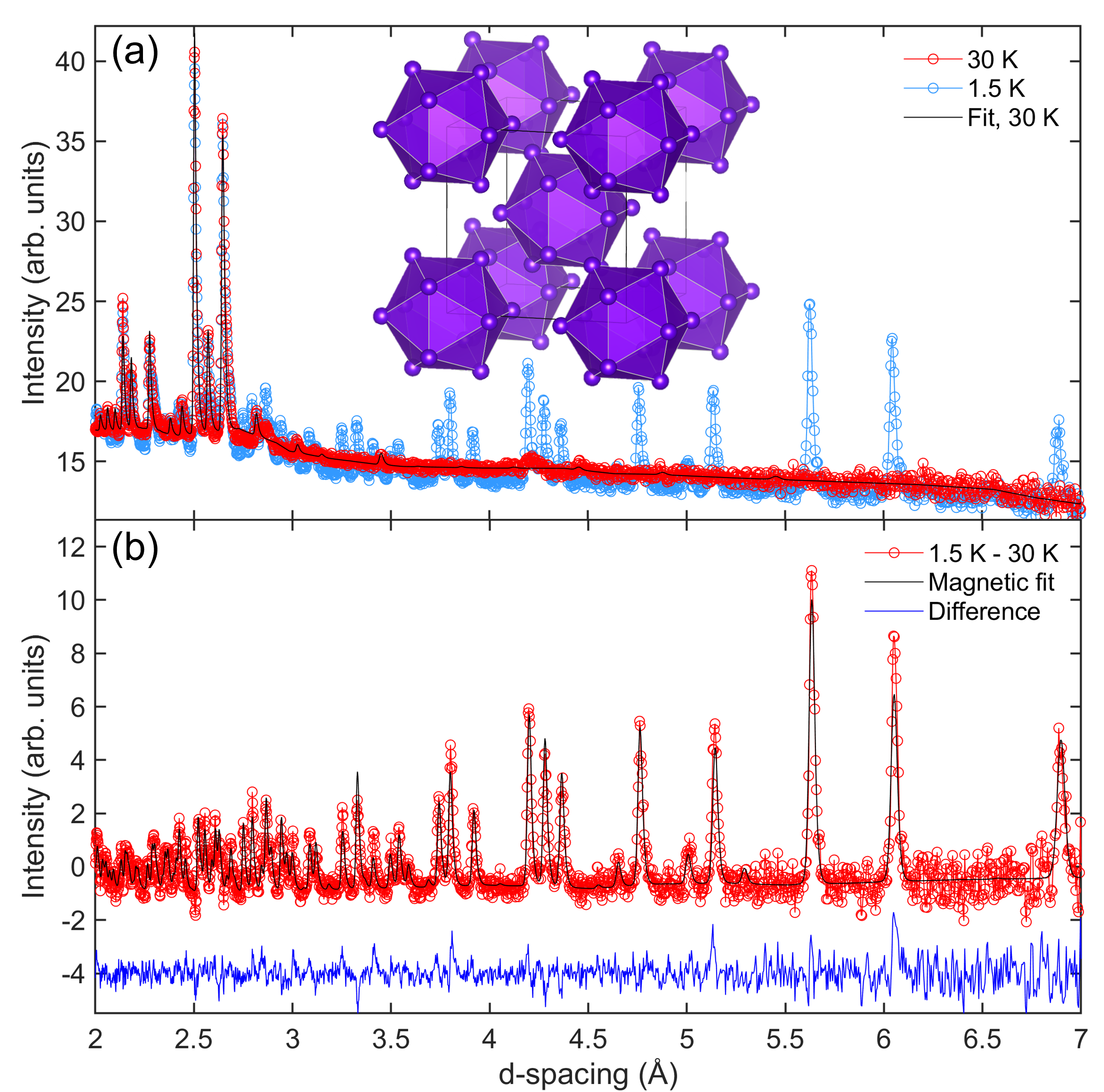}
    \caption{Powder neutron diffraction data on \ch{Cd6Tb}. (a) Data at 30~K, 1.5~K and the fit to the 30 K data. Inset: The positions of \ch{Tb^{3+}} ions in \ch{Cd6Tb}, with the icosahedral arrangement highlighted. (b) Magnetic diffraction pattern obtained by taking the difference between the 1.5~K and 30~K data in (a), together with the magnetic refinement (solid black line). The difference curve (blue) has been offset by 4 units for clarity.
    }
    \label{fig:data}
\end{figure}

A central challenge in this emerging field is to disentangle the roles of quasiperiodic geometry, chemical disorder, and random frustration in stabilizing unconventional magnetic states. Quasicrystal approximants (QCAs) \cite{Goldman1993} provide a crucial route to this goal by preserving the essential local geometry and connectivity of quasicrystals within a periodic framework [see the inset of Fig.~\ref{fig:data}(a) for the Tsai-type 1/1 QCA lattice], while remaining amenable to detailed microscopic probes. Various exotic magnetic structures arising from the QCA geometry have indeed been realized, such as the whirling noncoplanar spin textures which have recently been found in several Au-based QCAs \cite{Sato2019,Thilakan2024,Labib2024a}. They therefore offer a unique setting in which the intrinsic consequences of quasicrystal-derived geometry can be isolated and examined.

Among such systems, the Tsai-type 1/1 QCA \ch{Cd6Tb} occupies a unique position. It was the first quasicrystal-related material in which long-range magnetic order was experimentally established \cite{Tamura2010}. Moreover, since it is a binary compound, it naturally exhibits significantly lower levels of chemical disorder and site-occupancy randomness than its ternary counterparts. This makes it an exceptionally clean platform for addressing geometry-driven magnetism. Despite more than a decade of sustained interest, however, its microscopic magnetic structure has remained unresolved. Progress has been hindered by the large neutron absorption cross section of cadmium and by a subtle low-temperature cubic-to-monoclinic structural distortion \cite{Mori2012,Nishimoto2013,Tamura2002} that generates multiple inequivalent Tb environments.

In this paper, we overcome these obstacles and determine the magnetic structure of \ch{Cd6Tb} for the first time by combining high signal-to-noise powder neutron diffraction with rigorous symmetry-based analysis. We uncover a highly unusual noncoplanar magnetic ground state characterized by the coexistence of two distinct propagation vectors, resulting in a multi-\textit{k} magnetic structure. Importantly, the two propagation vectors are carried by different subsets of Tb moments.
An even more striking finding is that only one third of the Tb moments develop their full expected ordered value, while the remaining moments exhibit a strongly reduced static component. Muon-spin-rotation ($\mu$SR) measurements suggest that this reduction 
reflects persistent spin dynamics coexisting with long-range magnetic order. This coexistence points to magnetic fragmentation, in which the spin degrees of freedom separate into static and dynamic components.
Our results suggest \ch{Cd6Tb} as 
a candidate for the experimental realization of octahedral spin-ice physics.

In order to solve the magnetic structure of \ch{Cd6Tb}, we performed neutron powder diffraction (NPD) measurements at the cold neutron time-of-flight diffractometer WISH \cite{Chapon2011} at the ISIS facility in the UK. WISH offers a sufficiently low background that we were able to achieve a good signal-to-noise ratio even using natural cadmium. All refinements were carried out using FullProf \cite{RodriguezCarvajal1993}; further details of the sample synthesis, experimental methods and data analysis are available in the supplementary material \cite{supplementary}. 

Diffraction patterns at 30~K (above the magnetic ordering temperature) and at the base temperature of 1.5~K are shown in Fig.~\ref{fig:data}(a).
These data show that there are clear structural peaks at 30~K, which are well fitted by the high temperature cubic structural model; the monoclinic distortion is sufficiently subtle that we were unable to resolve it using our NPD data. Upon cooling to 1.5~K, a large number of additional strong Bragg peaks appear which we attribute to magnetic scattering. In order to simplify the problem, and because no monoclinic distortion was resolvable in the structural data, we worked in the high-temperature body-centered-cubic (bcc) unit cell which is the parent structure to both the structural and the magnetic distortion. Indexing the magnetic Bragg peaks relative to this unit cell revealed two propagation vectors, $\bm{k}_1=(100)$ and $\bm{k}_2=(\frac{1}{2}\frac{1}{2}0)$. Further, both these sets of peaks appeared at the same temperature despite at least two magnetic phase transitions being reported in the literature \cite{Tamura2010,Mori2012}, and in fact we were unable to resolve more than one magnetic phase transition within the resolution of the NPD data. 
\begin{figure}
    \centering
    \includegraphics[width=\linewidth]{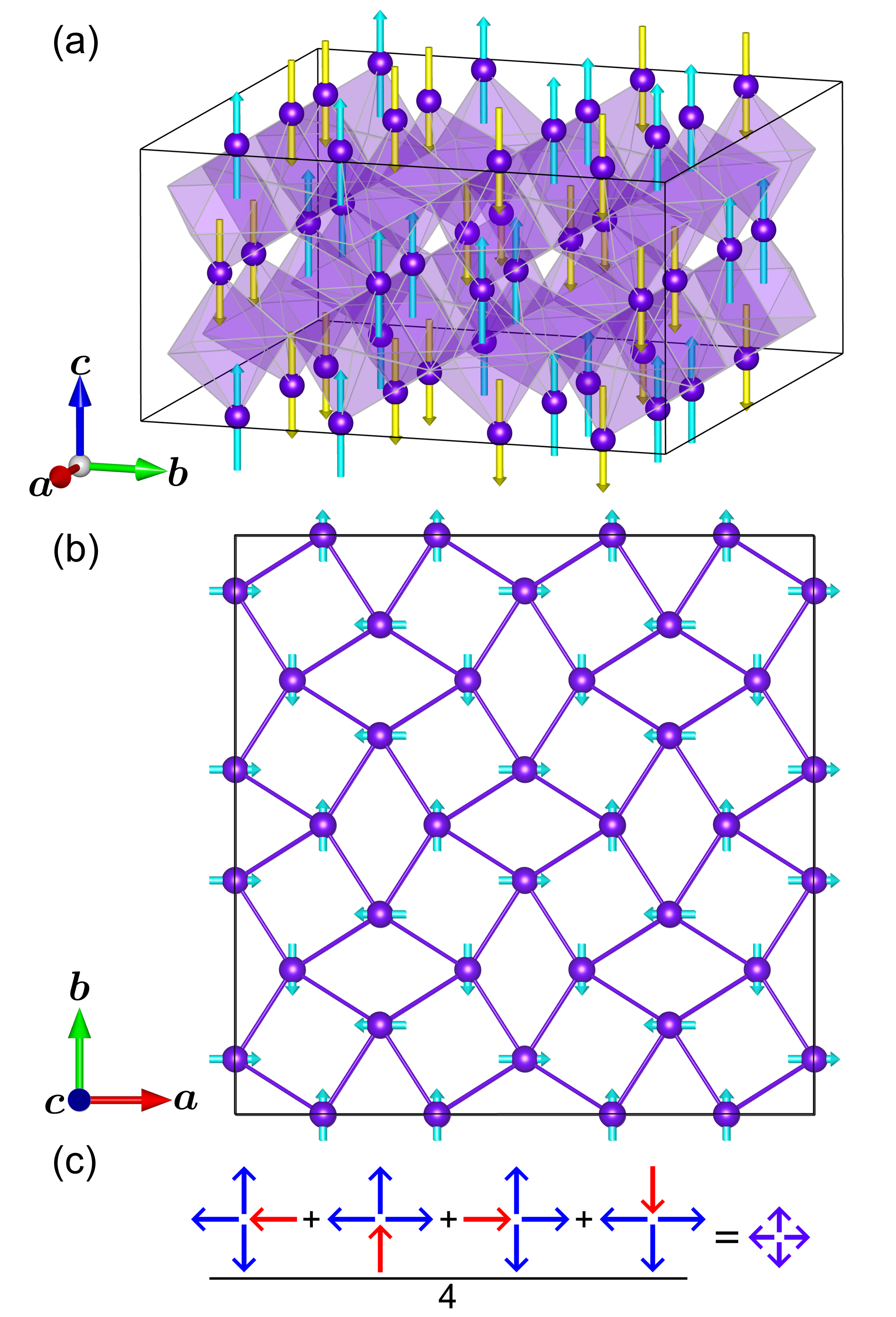}
    \caption{Magnetic structure of \ch{Cd6Tb}. (a,b) The refined magnetic structures of the apical ($z=0$, $\bm{k}_2$) spins (a) and equatorial ($\bm{k}_1$) spins (b) in \ch{Cd6Tb}. The black outlines in both panels indicate a $2\times2\times1$ magnetic supercell. In (a), down spins are highlighted in yellow for clarity. (c) The measured (average) moment can be half of the full moment if one randomly chosen spin in every four is flipped.}
    \label{fig:structure}
\end{figure}

Representational analysis \cite{Izyumov1991} was carried out separately for each of $\bm{k}_1$ and $\bm{k}_2$ using BasIReps \cite{RodriguezCarvajal1993}, again with respect to the high-temperature $Im\bar{3}$ space group. The sets of peaks corresponding to each propagation vector were fitted separately. In the high-temperature $Im\bar{3}$ phase, the \ch{Tb^{3+}} ions occupy a single Wyckoff position, 24g; the little group of the second propagation vector, $\bm{k}_2 = (\frac{1}{2}\frac{1}{2}0)$, splits the \ch{Tb^{3+}} sites into 4 orbits. It was found that by far the best fit to the $\bm{k}_2$ peaks was achieved using the one-dimensional $\mathrm{m}N_2^-$ representation on only two of the orbits, each of which have multiplicity 2. The corresponding \ch{Tb^{3+}} sites are ones with $z=0$ such that the only symmetry-allowed Ising direction is along $z$, and indeed it was found that these moments were ordered parallel to $z$, where the $xyz$ axes are defined to be parallel to the $abc$ axes of the cubic unit cell. This suggested that the spins in \ch{Cd6Tb} are likely to be Ising-like, as seen in other \ch{Tb^{3+}}-containing intermetallics \cite{Hiroto2020}. 

On the other hand, the violation of centering represented by the $\bm{k}_1$ propagation vector does not split the high temperature \ch{Tb^{3+}} Wyckoff position. 
Remarkably, we found that it was possible to fit the $\bm{k}_1$ peaks using only the orbits of $\bm{k}_2$ which were found to have no moment in the $\bm{k}_2$ structure, i.e., the different propagation vectors correspond to different sets of spins. In particular, by far the best fit to the $\bm{k}_1$ peaks was achieved with one of the high-symmetry directions of the $\mathrm{m}H_4^+$ representation. In the resulting structure, all sites have moments parallel to the only locally allowed Ising axis, leading to a non-coplanar structure. 
We note that there are two structures which can arise from the combination of signs of the magnetic modes corresponding to the two different propagation vectors. These structures have different local spin configurations, but it is not possible to distinguish them through diffraction. 
The resulting magnetic space group in both cases, $P 2_1/c.1^{\prime}_C[C 2/c]$ (\#14.84), corresponds to a structural space group and an allowed monoclinic distortion which are exactly those found in Refs.~\cite{Nishimoto2013,Tamura2005a, Liu2015}, lending further evidence that one of these is the correct structure. Indeed, $\bm{k}_1$ and $\bm{k}_2$ are coupled through this monoclinic distortion, as discussed in the supplementary material \cite{supplementary}. 

The refined non-coplanar, multi-$k$ magnetic structure is shown in Fig. \ref{fig:structure}(a,b), where the \ch{Tb^{3+}} ions are shown connected into a network of tilted corner-sharing octahedra rather than the isolated icosahedra shown in Fig.~\ref{fig:data}(a). In fact, the shortest Tb-Tb bonds are between icosahedra and within octahedra, as noted in \cite{Kim2012}, so this is a meaningful alternative perspective. In this picture, it is found that all the spins point either in or out of the octahedra. The $\bm{k}_1$ spins are aligned antiferromagnetically across octahedra and ferromagnetically with their other nearest neighbors with the same Ising axis, forming ferromagnetic layers. Their moments lie in the $ab$ plane, so we term these `equatorial spins'. Meanwhile, the $\bm{k}_2$ spins form chains, some of which are ferromagnetically aligned across octahedra and others which are antiferromagnetically aligned, with these chains then forming a checkerboard pattern in the $ab$-plane. These spins lie at the top and bottom of the octahedra as shown in Fig.~\ref{fig:structure}(a), so we term them `apical spins'. While these features are already surprising, perhaps the most surprising feature of the refined structure is that while the apical spins have the full ordered moment that would be expected for \ch{Tb^{3+}} ions [10.2(2)~$\mu_B$ compared to 9~$\mu_B$ for a free \ch{Tb^{3+}} ion], the equatorial spins have an ordered moment of about half of this, 4.62(9)~$\mu_B$. \footnote{We note that the listed uncertainties do not account for the uncertainty in the overall magnitude of the magnetic moments related to the strong neutron absorption in the sample, but that the \emph{ratio} of the moments on the different sites can be taken as reliable.} 

 \begin{figure}
\centering
     \includegraphics[width=\linewidth]{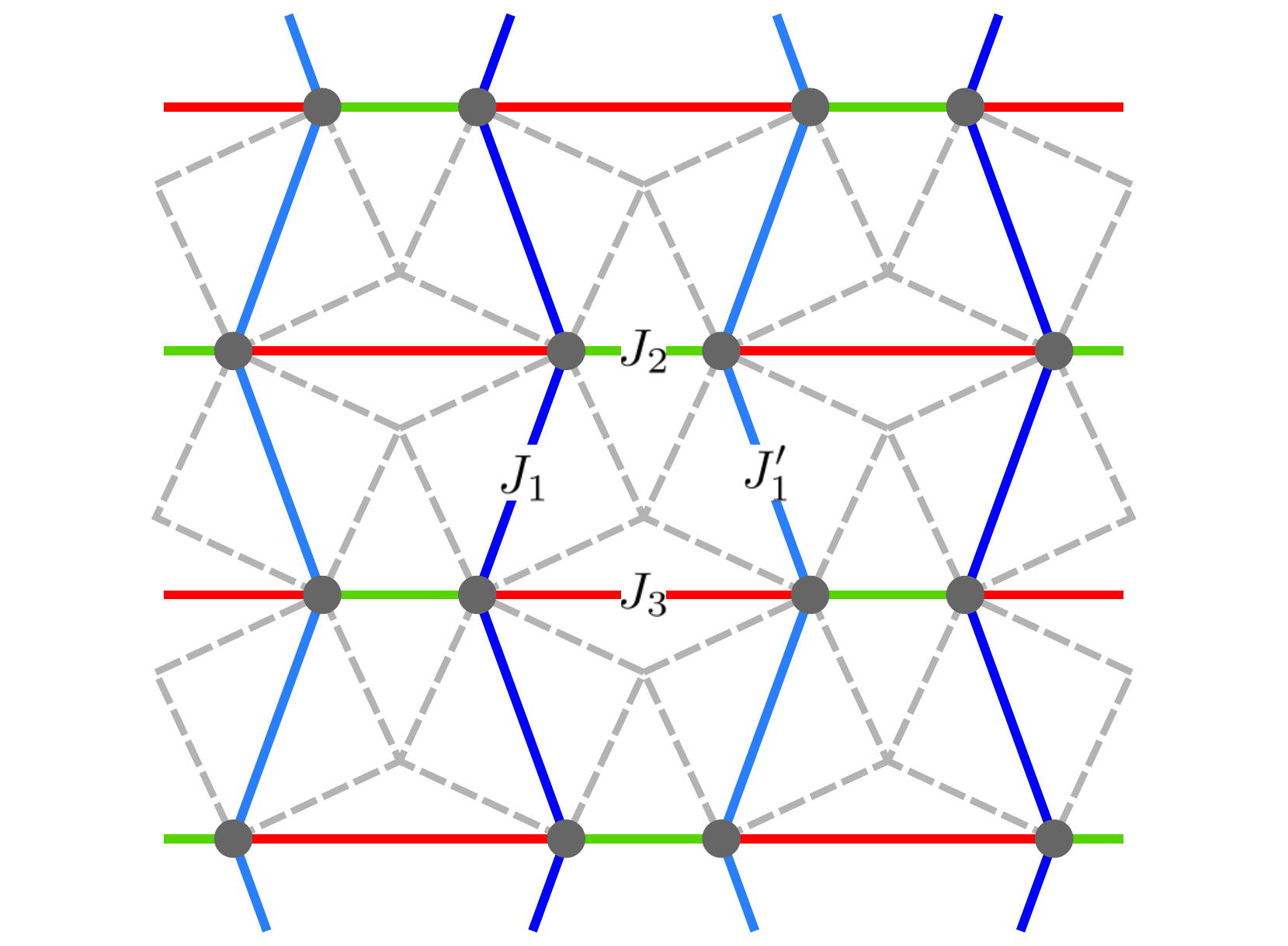}
     \caption{A two-dimensional version of the proposed model of frustrated Heisenberg exchange interactions. Gray dashed lines indicate the projection of the octahedra. Solid circles correspond to one of the three separate sets of spins sharing a common easy axis. $J_1^{\prime}$ is only different from $J_1$ for the apical spins.
     }
     \label{fig:interactions}
 \end{figure} 
 
The questions to be resolved are therefore: (i) what stabilizes this unusual ordering pattern, and (ii) why are some, but not all, of the ordered moments so reduced. 
To answer question (i), we propose a model of frustrated Heisenberg exchange interactions (see Fig.~\ref{fig:interactions} for a two-dimensional version and \cite{supplementary} for the three-dimensional model). These interactions do not couple spins with orthogonal easy axes, so this model allows us to consider the apical and equatorial spins separately. It is possible to reproduce both the $(100)$- and the $(\frac{1}{2}\frac{1}{2}0)$-type orders if $J_2$ is assumed to be ferromagnetic and $J_1$, $J_1^{\prime}$ and $J_3$ are all assumed to be antiferromagnetic. For the $(100)$-type ordering, $J_1 = J_1^{\prime}$, and these exchanges dominate over $J_3$. On the other hand, the $(\frac{1}{2}\frac{1}{2}0)$-type ordering is stabilized when $J_3$ dominates over $J_1,J_1^{\prime}$. This ordering requires $J_1 \neq J_1^{\prime}$ as otherwise there is a degenerate line of ground state propagation vectors $\bm{k}=(\frac{1}{2}\frac{1}{2}k_z)$; a difference between $J_1$ and $J_1^{\prime}$ for the apical spins is indeed symmetry-allowed in the low-temperature monoclinic unit cell.  If the parameters lie close to a phase boundary, it can be assumed that the slight symmetry breaking between the apical and equatorial sites could stabilize these different orders for the different sets of spins.
Monoclinic distortion thus plays a role in selecting an ordered state out of the degenerate manifold, and, more broadly, symmetry lowering may provide a general mechanism by which frustrated Tsai-type manifolds avoid spin freezing and instead stabilize ordered states.

A possible scenario that would answer question (ii) would be one in four of the equatorial spins being flipped at random from their ordered state; the average moment that would contribute to the Bragg peaks would then be half of the full moment, as illustrated in Fig.~\ref{fig:structure}(c). This scenario would mean that the equatorial spins \emph{fragment} into an ordered component, consisting of monopolar and quadrupolar configurations, and a disordered dipolar component. 
Such a fragmentation has been predicted to arise in octahedral spin ice \cite{Szabo2022}, a proposal which requires Ising moments at the vertices of corner-sharing octahedra interacting via dipole-dipole or staggered Dzyaloshinskii-Moriya (DM) interactions and which has not previously been experimentally realized. 
Like the celebrated tetrahedral spin ice model found in pyrochlores \cite{Bramwell2001,Fennell2009},
this model has a highly degenerate ground state manifold. However, unlike on a tetrahedron, it is possible to have symmetry-distinct monopole-free (3 in - 3 out) arrangements on an octahedron, leading to additional phenomena and in particular, the fragmentation of degrees of freedom into dipolar and quadrupolar components. Moreover, interactions between spins on different octahedra can lead to a variety of unusual multi-$k$ ground states \cite{Szabo2022}. We hypothesize that the additional complexity introduced by the tilting of the octahedra and the subtle monoclinic distortion could stabilize the observed structure. Alternatively, the physics may be more closely related to nematic phases, in which different combinations of monopolar, dipolar and quadrupolar configurations are energetically favored, which have very recently been predicted in other regions of the phase diagram for the same lattice geometry \cite{Stern2026}. While the temperature scale of the ordering ($\approx 20$~K) means that dipole-dipole interactions are too weak to play a significant role, staggered DM interactions arising from the Ruderman–Kittel–Kasuya–Yosida (RKKY) mechanism are likely to be strong enough to mediate the interaction between nearest-neighbor orthogonal spins.

\begin{figure}
    \centering
    \includegraphics[width=\linewidth]{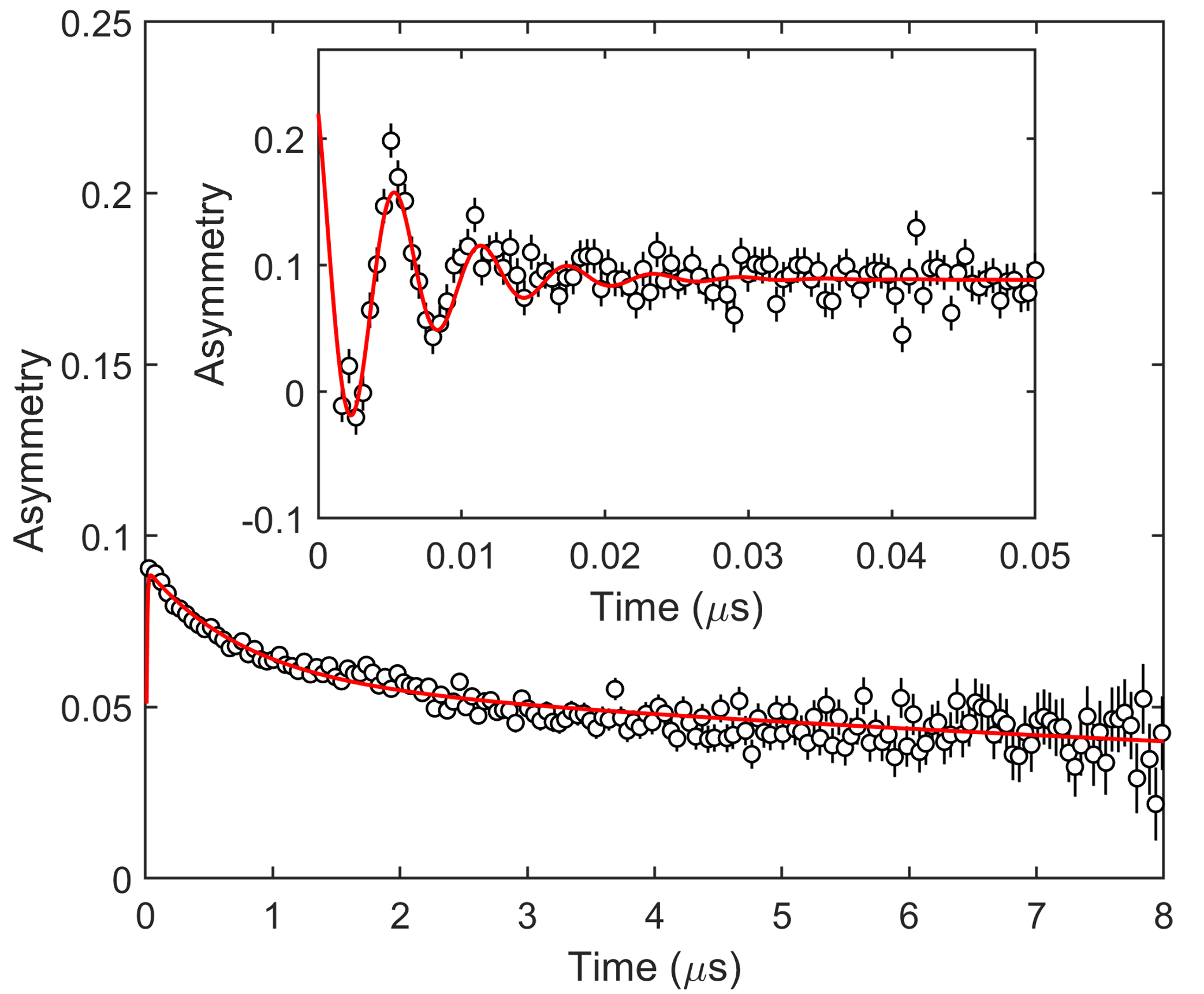}
    \caption{$\mu$SR data taken at 5 K. The main panel shows the long time decay of the asymmetry while the inset shows a zoomed in view of the short time oscillatory behaviour. Error bars correspond to one standard deviation statistical uncertainty.}
    \label{fig:muons}
\end{figure}

The suggestion of a disordered dipolar component to the spins raises the further question of whether this disorder is static or dynamic. To address this, we performed $\mu$SR measurements on \ch{Cd6Tb} on the GPS instrument~\cite{GPS} at the Paul Scherrer Institute in Switzerland. 
The results at 5~K are shown in Fig.~\ref{fig:muons}. At very short times (inset), strongly damped high frequency oscillations are seen, indicating long-range static magnetic order. The frequency of these oscillations corresponds to a magnetic field of 1.22(4)~T.
At long times however, the asymmetry decays to well below the 1/3 tail level expected in the presence of exclusively static magnetism, and continues to decay slowly throughout the available time window. The slowly decaying tail can be fitted with a sum of two exponentials, with decay rates 1.3(2) and 0.059(17) $\mu$s$^{-1}$, respectively, 
with the slower decay rate corresponding to fluctuations on time scales of 10~$\mu$s. These data therefore indicate that while there is static magnetism on timescales of at least tens of nanoseconds, consistent with the presence of peaks in neutron diffraction, there are also dynamics on slower time scales, which are probed by $\mu$SR measurements but invisible in neutron diffraction. These findings are consistent with the picture of spin reduction due to dynamic spin fragmentation of some but not all of the spins. Intriguingly, a similar $\mu$SR signal with coexisting high-frequency oscillations and long-time decay has recently been observed in a Tb-based frustrated kagom\'{e} material \cite{Devi2026}. While 5~K was the lowest temperature at which we were able to obtain reliable $\mu$SR data, in the supplementary material \cite{supplementary}, we provide evidence from additional $\mu$SR measurements and from AC susceptibility measurements that the dynamical behavior in \ch{Cd6Tb} persists down to 1.5~K, despite the transition at 2.4~K \cite{Tamura2010}.  

In summary, we have determined for the first time the magnetic structure of the prototypical magnetic QCA, \ch{Cd6Tb}. We have found a highly unusual non-coplanar, multi-$k$ magnetic ground state, in which some but not all of the ordered moments are significantly reduced from their expected full value. This moment reduction is attributed to persistent fluctuations of the spins coexisting with static magnetic order, as evidenced by $\mu$SR. The combination of the magnetic topology, consisting of Ising spins at the vertices of corner-sharing octahedra, and the unusual magnetic behavior makes this material a candidate for the first experimental realization of octahedral spin-ice physics. This, in turn, has a significant impact on the many magnetic quasicrystal-related materials with a similar arrangement of magnetic ions which have been found to form spin-glass-like states at low temperatures~\cite{Ibuka2011,Kashimoto2006,Labib2022}.  
Our findings suggest that these systems may instead host 
highly correlated spin-\emph{ices} 
rather than conventional random spin-glass states. Indeed, unusual relaxation behaviours, unconventional excitations, and the presence of short-range magnetic correlations have repeatedly indicated that the low-temperature phases of magnetic quasicrystals and approximants cannot be fully understood within the framework of conventional spin glasses~\cite{AlQadi2009,Sato2006,Sato2002,Ibuka2011}. 
What is perhaps even more striking is that this unconventional magnetic state emerges at temperatures of order 20~K, far above the millikelvin energy scale characteristic of dipole-dipole coupled spin-ice systems. Our findings therefore highlight a potential pathway toward stabilizing exotic frustrated magnetic states at elevated temperatures.

\begin{acknowledgments}
\emph{Acknowledgments} --- We acknowledge helpful discussions with A. Szab\'{o}. L.W. acknowledges support from a grant from the UKRI International Science Partnerships Fund (award ISPF-229) for partnership development between ISIS, Diamond and the Paul Scherrer Institute. This work was also supported by the Japan Society for the Promotion of Science through Grants-in-Aid for Scientific Research (Grants No. JP19H05817, No. JP19H05818, No. JP21H01044, and No. JP24K17016) and the Japan Science and Technology agency, CREST, Japan, through a grant No. JPMJCR22O3. 
The neutron scattering experiment at the ISIS Neutron and Muon Source was supported by beamtime allocation RB2420089 from the Science and Technology Facilities Council~\cite{wish_doi} . Data are available at \url{https://doi.org/10.5286/ISIS.E.RB2420089}. This work is based on experiments performed at the Swiss Muon Source SµS, Paul Scherrer Institute, Villigen, Switzerland. We would like to thank G. Stenning for help with the AC susceptibility measurements using the MPMS instrument in the Materials Characterisation Laboratory at the ISIS Neutron and Muon Source. Some figures were made using \textsc{Vesta}~\cite{Momma2011}.
\end{acknowledgments}

\nocite{Arnold2014,Gomez2003,Tsai2000,Tsai2008a, Isotropy, Campbell2006, Suter2012}

\newpage
\section*{Methods}

\paragraph{Sample synthesis}
A polycrystalline alloy with a nominal composition of Cd$_6$Tb was synthesized by melting high-purity constituent elements in an alumina crucible sealed in an evacuated quartz tube under an Ar atmosphere at 973~K for 10~h. The obtained sample was subsequently annealed at 773~K for 120~h to improve chemical homogeneity and reduce point defects toward their thermal equilibrium concentration. 
Phase purity of the sample was verified by powder X-ray diffraction (XRD).

\paragraph{Neutron diffraction}
Powder neutron diffraction was performed on the WISH diffractometer at ISIS \cite{Chapon2011}. 1.5~g of sample was loaded into a thin-walled 3~mm vanadium can. 
Measurements with long counting times of 233 and 218~$\mu$Ah (typical current of 40~$\mu$A of protons on target) were taken at 1.5 and 30~K, respectively, to measure the diffraction patterns in both the base-temperature ordered phase and in the paramagnetic phase. Additional measurements were taken at a series of intermediate temperatures to probe the effect of the successive phase transitions.
Data were reduced using \textsc{Mantid} \cite{Arnold2014}. 
For the structural refinement, data from detector banks with average scattering angles $2\theta$ of 90.0$^{\circ}$ and 121.66$^{\circ}$ were used, as these banks were the least affected by the strong neutron absorption. 
For the magnetic refinement, only the detector bank corresponding to $2\theta=90^{\circ}$ was used.
Magnetic structure solution was performed using FullProf and BasIReps \cite{RodriguezCarvajal1993}. Symmetry analysis was performed using ISODISTORT \cite{Isotropy,Campbell2006}.

\paragraph{Muon spin rotation}
$\mu$SR was performed on the GPS instrument at PSI \cite{GPS}. A 0.5~g powder sample of \ch{Cd6Tb} was enclosed between two pieces of Kapton tape. The sample was placed in a helium flow cryostat with a base temperature of 1.5~K and measurements were taken at a series of temperatures including 5~K. The measurement at 300~K was used for calibration. The data were analysed using \cite{Suter2012}.

\paragraph{AC susceptibility}
AC magnetic susceptibility measurements were performed using a Quantum Design Magnetic Property Measurement System 3 on a 41.9 mg powder sample of \ch{Cd6Tb}. Measurements were performed at several frequencies.

\end{document}

\title{Supplementary material: Magnetic ground state of a prototype quasicrystal approximant: a candidate for octahedral spin ice physics}
\author{Leonie Woodland}
\thanks{These authors contributed equally to this work.}
\affiliation{ISIS Facility, Rutherford Appleton Laboratory, Chilton, Didcot OX11 0QX, UK}
\affiliation{Clarendon Laboratory, University of Oxford Physics Department, Parks Road, Oxford OX1 3PU, UK}
\author{Farid Labib}
\thanks{These authors contributed equally to this work.}
\affiliation{Research Institute for Science and Technology, Tokyo University of Science, Tokyo 125-8585, Japan}
\author{Dmitry Khalyavin}
\affiliation{ISIS Facility, Rutherford Appleton Laboratory, Chilton, Didcot OX11 0QX, UK}
\author{Pascal Manuel}
\affiliation{ISIS Facility, Rutherford Appleton Laboratory, Chilton, Didcot OX11 0QX, UK}
\author{Fabio Orlandi}
\affiliation{ISIS Facility, Rutherford Appleton Laboratory, Chilton, Didcot OX11 0QX, UK}
\author{Hubertus Luetkens}
\affiliation{Laboratory for Muon-Spin Spectroscopy, Paul Scherrer Institut, CH-5232 Villigen PSI, Switzerland}
\author{Ryuji Tamura}
\affiliation{Department of Materials Science and Technology, Tokyo University of Science, Tokyo 125-8585, Japan}

\maketitle

\section{Sample synthesis}
A polycrystalline alloy with a nominal composition of Cd$_6$Tb was synthesized by melting high-purity constituent elements in an alumina crucible sealed in an evacuated quartz tube under an Ar atmosphere at 973~K for 10~h. The obtained sample was subsequently annealed at 773~K for 120~h to improve chemical homogeneity and reduce point defects toward their thermal equilibrium concentration. 
Phase purity of the sample was verified by powder X-ray diffraction (XRD).

\section{Neutron diffraction: experimental details}

Powder neutron diffraction was performed on the WISH diffractometer at ISIS \cite{Chapon2011}. 1.5~g of sample was loaded into a thin-walled 3~mm vanadium can. 
Measurements with long counting times of 233 and 218~$\mu$Ah (typical current of 40~$\mu$A of protons on target) were taken at 1.5 and 30~K, respectively, to measure the diffraction patterns in both the base-temperature ordered phase and in the paramagnetic phase. Additional measurements were taken at a series of intermediate temperatures to probe the effect of the successive phase transitions.
Data were reduced using \textsc{Mantid}~\cite{Arnold2014}. 
For the structural refinement, data from detector banks with average scattering angles $2\theta$ of 90.0$^{\circ}$ and 121.66$^{\circ}$ were used, as these banks were the least affected by the strong neutron absorption. 
For the magnetic refinement, only the detector bank corresponding to $2\theta=90^{\circ}$ was used.

\section{Nuclear structure}\label{sec:parent}

The high-temperature parent cubic $Im\bar{3}$ structure consists of Tsai-type clusters arranged in a body-centered cubic arrangement \cite{Gomez2003,Tsai2000}. The Tsai-type cluster in \ch{Cd6Tb} consists of four layers going out from the center: an inner tetrahedron of Cd, then a dodecahedron of Cd, then an icosahedron of Tb and finally an icosidodecahedron of Cd \cite{Tsai2008a}. In the high temperature phase, the inner tetrahedron is orientationally disordered such that the average structure can be represented by an icosahedron where each of the sites has 1/3 occupancy. The monoclinic structural transition is driven by orientational ordering of the tetrahedra \cite{Nishimoto2013}.

The structural refinement was performed with respect to the high temperature cubic structure as the subtle monoclinic distortion could not be resolved in our powder neutron diffraction data. The atomic positions were then fixed when performing the magnetic refinement. The results of the structural refinement are given in Table~\ref{t:nuclear_struct}.

\setlength{\extrarowheight}{2pt}
\begin{longtable}{l l}
    \caption{The nuclear crystal structure, corresponding to the refinement shown in Fig.~1(a) of the main text.  The Bragg R-factor for the bank at $2\theta=121.66^{\circ}$ is 18.0\%. \label{t:nuclear_struct}}\\
\hline
    \hline
    \endfirsthead
    \caption{(continued)}\\
    \hline
    \hline
    \endhead
    \hline
    \hline
    \endfoot
    \hline
    \hline
    \endlastfoot
    Space group & $Im\bar{3}$ \\
    Space group number & 204 \\
    Point group & $m\bar{3}$ \\
Unit cell parameter (\AA) & \makecell[tl]{$a=15.4144(9)$}\\
    Space group symmetry operations (24) & $x,y,z$\\
&$x,-y,-z$\\
&$-x,y,-z$\\
&$-x,-y,z$\\
&$y,z,x$\\
&$-y,-z,x$\\
&$y,-z,-x$\\
&$-y,z,-x$\\
&$z,x,y$\\
&$-z,x,-y$\\
&$-z,-x,y$\\
&$z,-x,-y$\\
&$-x,-y,-z$\\
&$-x,y,z$\\
&$x,-y,z$\\
&$x,y,-z$\\
&$-y,-z,-x$\\
&$y,z,-x$\\
&$-y,z,x$\\
&$y,-z,x$\\
&$-z,-x,-y$\\
&$z,-x,y$\\
&$z,x,-y$\\
&$-z,x,y$\\
Space group centering (2) & $x,y,z$ \\
     & $x+\frac{1}{2}, y+\frac{1}{2},z+\frac{1}{2}$ \\
     \makecell[tl]{Fractional coordinates, $U_{\text{iso}}$ \\ and occupancy for each atom} & \makecell[tl]{Cd1 0.1182(6) 0.3383(5) 0.2024(7) 0.014(3) 1.0 \\
     Cd2 0.0 0.059(3) 0.103(3) 0.014(3) 0.333 \\
     Cd3 0.0 0.2391(8) 0.0956(9) 0.014(3) 1.0 \\
     Cd4 0.0 0.4068(10) 0.3478(10) 0.014(3) 1.0 \\
     Cd5 0.1548(5) 0.1548(5) 0.1548(5) 0.014(3) 1.0 \\
     Cd6 0.1952(10) 0.0 0.5 0.014(3) 1.0 \\
     Cd7 0.4049(15) 0.0 0.0 0.014(3) 1.0 \\
     Tb 0.0 0.1915(6) 0.3029(5) 0.010(3) 1.0\\}\\
    \end{longtable}

\section{Magnetic structure solution}

\subsection{Representational analysis relative to $Im\bar{3}$}
\subsubsection{\hho propagation vector}
As discussed in the main text, the $\bm{k}_2=$ \hho propagation vector splits the single Tb Wyckoff position in the parent $Im\bar{3}$ structure into 4 orbits, with multiplicities 4, 4, 2 and 2, labeled as Tb1, Tb2, Tb3 and Tb4 respectively. A good fit was obtained using the one-dimensional $\mathrm{m}N_2^-$ representation on only orbits Tb3 and Tb4 (we follow the ISODISTORT \cite{Isotropy, Campbell2006} notation for irreducible representations and order parameter directions). The magnitude of the moments on both orbits was constrained to be the same since, when they were allowed to refine independently, very similar values were obtained for each.

\subsubsection{$(100)$ propagation vector}
It was found that the peaks indexed by propagation vector $\bm{k}_1=(100)$ were best fitted with a structure corresponding to the $\mathrm{m}{H}_4^+$ representation. This is a 3-dimensional representation, and there are 5 copies contained within the magnetic representation of $Im\bar{3}$. 
The solution was searched for along special directions of the order parameter, which are most likely, because phases with non-special order parameter directions are less symmetric and require significant contributions from high-power invariants in the Landau free energy decomposition. Indeed, it was found that a good fit was obtained using the $(a00)$ direction of the order parameter.
Further, the refined structure had non-negligible moments corresponding only to one copy of the representation, leaving only one free parameter. The resulting structure corresponded to non-zero moments on only the Tb1 (or Tb2) orbit of the little group of $(\frac{1}{2}\frac{1}{2}0)$.
We therefore assumed that \emph{both} the Tb1 and the Tb2 orbits of the little group of $(\frac{1}{2}\frac{1}{2}0)$ order according to the $(100)$ propagation vector, with the basis vectors transformed appropriately. The magnitude of the moments on both sites was constrained to be the same because the diffraction pattern cannot distinguish between moments on only one site, or on both, so the most symmetric option was chosen. Whether the relative phase between the Tb1 and Tb2 orbits is $+1$ or $-1$ produces different structures which cannot be distinguished by diffraction, as discussed below.

\subsection{Possible magnetic structures}

\begin{figure}
    \centering
    \includegraphics[width=\linewidth]{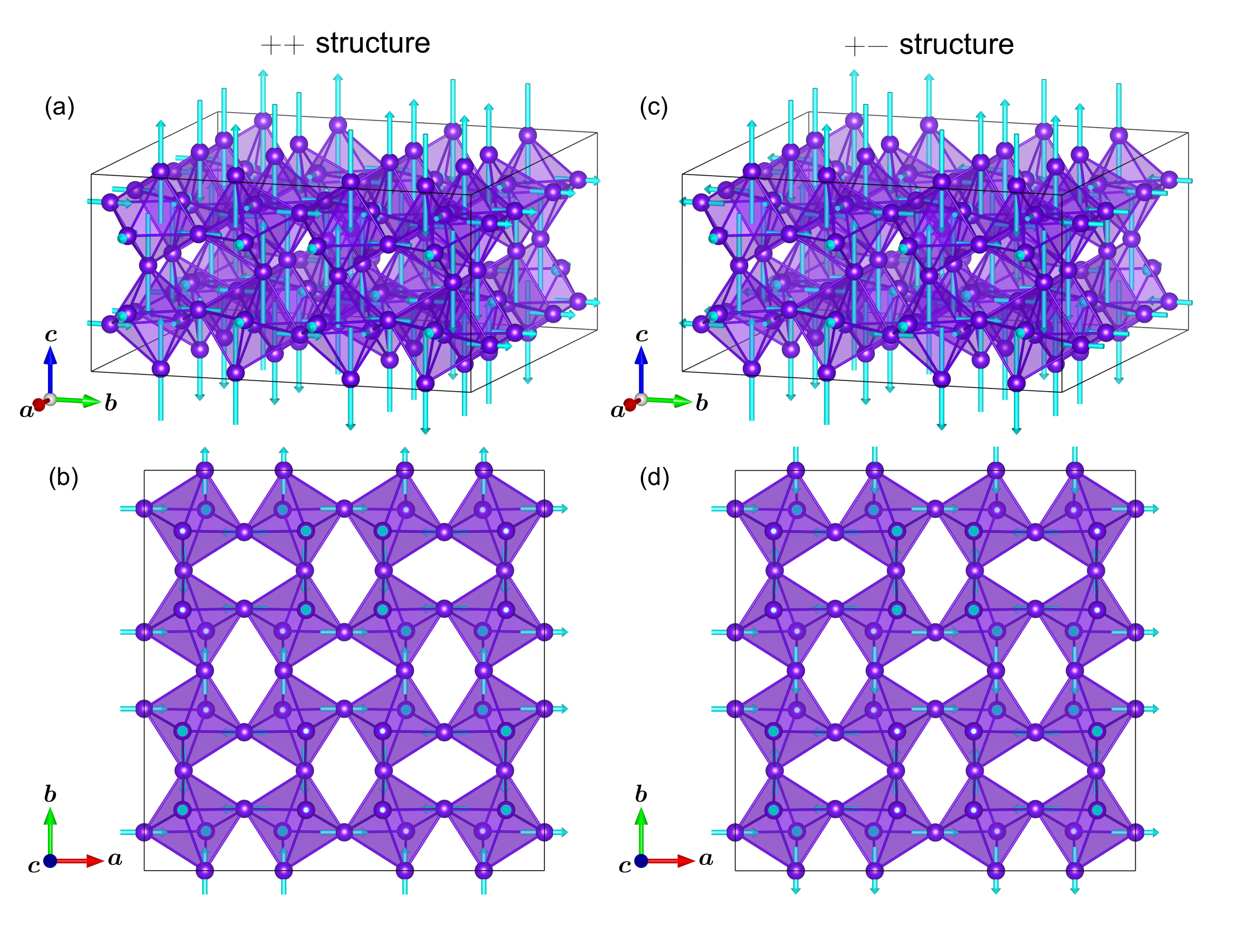}
    \caption{The two distinct ways of combining the structures for the apical and equatorial spins together. The $+-$ structure (c,d) is related to the $++$ structure (a,b) by flipping the spins with moment parallel to the $b$-direction of the high-temperature structural cubic unit cell. The solid black lines in all panels indicate a $2\times2\times1$ magnetic supercell of the cubic parent. In (b,d), larger circles correspond to spins pointing out of the page whereas smaller circles correspond to spins pointing into the page.}
    \label{fig:structureoptions}
\end{figure}

As mentioned in the main text, there are two possibilities for how the structures of the equatorial and apical spins are combined. This corresponds to an ambiguity in the relative phase of the Fourier coefficients of the Tb1 and Tb2 type ions, which it is not possible to resolve in a diffraction experiment.  The structure shown in Fig.~\ref{fig:structureoptions}(a,b) corresponds to the structure in which the phases of both the Tb1 and Tb2 spins are +1, labelled here as the ``++" structure. Table~\ref{t:ppft} describes this structure in terms of Fourier coefficients of the two propagation vectors for each of the Tb ions in the primitive unit cell of the parent $Im\bar{3}$ structure, while Table~\ref{t:ppmsg} describes this magnetic structure in terms of the resulting magnetic space group.

    \begin{longtable}{l c c c}
    \caption{The ++ magnetic structure, shown in terms of Fourier coefficients for each Tb ion in the primitive unit cell of the parent $Im\bar{3}$ structure. The corresponding magnetic R-factor is $R_{\text{mag}}=11.6\%$. \label{t:ppft}}\\
\hline
    \hline
    \endfirsthead
    \caption{(continued)}\\
    \hline
    \hline
    \endhead
    \hline
    \hline
    \endfoot
    \hline
    \hline
    \endlastfoot
    Atom$(x,y,z)$/propagation vector & $M_x$ ($\mu_{\rm{B}}$) & $M_y$ ($\mu_{\rm{B}}$) & $M_z$ ($\mu_{\rm{B}}$) \\
    \hline
     Tb1\_1(0,0.1915,0.3029)/$\bm{k}_1$    & 4.62(9) & 0 & 0 \\
     Tb1\_2(0,-0.1915,0.3029)/$\bm{k}_1$ & 4.62(9) & 0 & 0 \\
     Tb1\_3(0,-0.1915,-0.3029)/$\bm{k}_1$ & 4.62(9) & 0 & 0 \\
     Tb1\_4(0,0.1915,-0.3029)/$\bm{k}_1$ & 4.62(9) & 0 & 0 \\
     \hline
     Tb2\_1(0.3029,0,0.1915)/$\bm{k}_1$ & 0 & 4.62(9) & 0 \\
     Tb2\_2(-0.3029,0,0.1915)/$\bm{k}_1$ & 0 & 4.62(9) & 0 \\
     Tb2\_3(-0.3029,0,-0.1915)/$\bm{k}_1$ & 0 & 4.62(9) & 0 \\
     Tb2\_4(0.3029,0,-0.1915)/$\bm{k}_1$ & 0 & 4.62(9) & 0 \\
     \hline
     Tb3\_1(0.1915,0.3029,0)/$\bm{k}_2$ & 0 & 0 & 10.2(2) \\
     Tb3\_2(-0.1915,-0.3029,0)/$\bm{k}_2$ & 0 & 0 & -10.2(2) \\
     \hline
     Tb4\_1(-0.1915,0.3029,0)/$\bm{k}_2$ & 0 & 0 & 10.2(2) \\
     Tb4\_2(0.1915,-0.3029,0)/$\bm{k}_2$ & 0 & 0 & -10.2(2)\\
    \end{longtable}

    \begin{longtable}{l  l}
        \caption{The ++ structure, described in terms of its magnetic space group. The atomic positions are fixed to those from the 30~K refinement. Note that the Tb site labels are unrelated to those used in Table~\ref{t:ppft} and that the standard setting for the magnetic space group is used, so that the definitions of the axes and propagation vectors do not correspond to those used in the main text and in the representational analysis above.
        \label{t:ppmsg}}
        \\
    \hline
    \hline
    \endfirsthead
    \caption{(continued)}\\
    \hline
    \hline
    \endhead
    \hline
    \hline
    \endfoot
    \hline
    \hline
    \endlastfoot
    Parent space group & $Im\bar{3}.1^{\prime}$ \\
    Propagation vector(s) & $(100)$, $(\frac{1}{2},0,\frac{1}{2})$ \\
    \makecell[tl]{Transformation from parent basis to \\ the one used} & $a+c,b,-a+c;0,0,\frac{1}{2}$ \\
    MSG symbol (BNS) & $P_C 2_1/c$ \\
    MSG number (BNS) & 14.84 \\
    MSG symbol (UNI) & $P 2_1/c.1^{\prime}_C[C 2/c]$ \\
    \makecell[tl]{Transformation from basis used to \\standard setting of MSG} & $a,b,c;0,0,0$ \\
    Magnetic point group & $2/m.1^{\prime}$ \\
    Active irreps & $\mathrm{m}{N}_2^-$, primary order parameter \\
    & $\mathrm{m}{H}_4^+$, special direction, primary order parameter \\
    Unit cell parameters (\AA, $^{\circ}$) & \makecell[tl]{$a=21.91182, b=15.49400, c=21.91182,$ \\ $\alpha=90.0, \beta=90.0, \gamma=90.0$}\\
    MSG symmetry operations (4) & $x,y,z,+1$ \\
    & $-x, y+\frac{1}{2}, -z+\frac{1}{2},+1$ \\
    & $-x, -y, -z, +1$ \\
    & $x, -y+\frac{1}{2}, z+\frac{1}{2}, +1$\\
    MSG symmetry centering operations (2) & $x,y,z,+1$ \\
     & $x+\frac{1}{2}, y+\frac{1}{2},z,-1$ \\
     Positions and occupancies of magnetic atoms & Tb1 Tb 0.34575 0.80288 0.65425 1.0\\
     & Tb2 Tb 0.84575 0.80288 0.15425 1.0\\
     & Tb3 Tb 0.40144 0.69149 -0.09856 1.0\\
     & Tb4 Tb -0.09856 0.69149 0.40144 1.0\\
     & Tb5 Tb 0.49719 0.50000 0.69431 1.0\\
     & Tb6 Tb 0.80569 0.50000 0.00281 1.0\\
     Positions and occupancies of non-magnetic atoms & Cd1 Cd 0.47822 0.70237 0.63996 1.0\\
     & Cd2 Cd -0.02178  0.70237  0.13996 1.0 \\
     & Cd3 Cd  0.63996  0.29763  0.47822 1.0 \\
     & Cd4  Cd 0.13996  0.29763 -0.02178 1.0\\
    & Cd5  Cd  0.52031  0.61819  0.81795 1.0\\
    & Cd6  Cd  0.02031  0.61819  0.31795 1.0\\
    & Cd7  Cd  0.81795  0.38181  0.52031 1.0\\
    & Cd8  Cd  0.31795  0.38181  0.02031 1.0\\
    & Cd9  Cd  0.41027  0.83826  0.79209 1.0\\
    & Cd10 Cd -0.08973  0.83826  0.29209 1.0\\
    & Cd11 Cd  0.70791  0.16174  0.08973 1.0\\
    & Cd12 Cd  0.20791  0.16174  0.58973 1.0\\
    & Cd13 Cd  0.27956  0.60280  0.72044 0.333\\
    & Cd14 Cd  0.77956  0.60280  0.22044 0.333\\
    & Cd15 Cd  0.33096  0.50000  0.72816 0.333\\
    & Cd16 Cd  0.72816  0.50000  0.33096 0.333\\
    & Cd17 Cd  0.30140  0.55913  0.80140 0.333\\
    & Cd18 Cd  0.80140  0.55913  0.30140 0.333\\
    & Cd19 Cd  0.36954  0.59558  0.63046 1.0\\
    & Cd20 Cd  0.86954  0.59558  0.13046 1.0\\
    & Cd21 Cd  0.41733  0.50000  0.82175 1.0\\
    & Cd22 Cd  0.82175  0.50000  0.41733 1.0\\
    & Cd23 Cd  0.29779  0.73909  0.79779 1.0\\
    & Cd24 Cd  0.79779  0.73909  0.29779 1.0\\
    & Cd25 Cd  0.45337  0.84783  0.54663 1.0\\
    & Cd26 Cd -0.04663  0.84783  0.04663 1.0\\
    & Cd27 Cd  0.62728  0.50000  0.77946 1.0\\
    & Cd28 Cd  0.77946  0.50000  0.62728 1.0\\
    & Cd29 Cd  0.42391 -0.09325 -0.07609 1.0\\
    & Cd30 Cd -0.07609 -0.09325  0.42391 1.0\\
    & Cd31 Cd  0.40482  0.65482  0.75000 1.0\\
    & Cd32 Cd -0.09518  0.65482  0.25000 1.0\\
    & Cd33 Cd  0.75000  0.34518  0.40482 1.0\\
    & Cd34 Cd  0.25000  0.34518 -0.09518 1.0\\
    & Cd35 Cd  0.34762  0.00000  0.84762 1.0\\
    & Cd36 Cd  0.50000  0.69525  0.50000 1.0\\
    & Cd37 Cd  0.59762  0.50000 -0.09762 1.0\\
    & Cd38 Cd  0.45243  0.50000 -0.04757 1.0\\
    & Cd39 Cd  0.75000  0.09514  0.25000 1.0\\
    & Cd40 Cd  0.75000 -0.09514  0.25000 1.0\\
    & Cd41 Cd  0.45243  0.50000  0.54757 1.0\\
    \makecell[tl]{Magnetic moment components ($\mu_{\rm{B}}$) \\
    of magnetic atoms, symmetry \\ 
    constraints and moment magnitudes} &
    \makecell[tl]{
    Tb1   3.27  0.00000   3.27 Mx,My,Mz 4.62(9) \\
Tb2   3.27  0.00000   3.27 Mx,My,Mz 4.62(9) \\
Tb3   3.27  0.00000  -3.27 Mx,My,Mz 4.62(9) \\
Tb4   3.27  0.00000  -3.27 Mx,My,Mz 4.62(9) \\
Tb5   0.00000  10.2(2)  0.00000 Mx,My,Mz 10.2(2) \\
Tb6   0.00000 -10.2(2)  0.00000 Mx,My,Mz 10.2(2)}
 \end{longtable}

The other possibility for the magnetic structure, which we label as ``+--" is shown in Fig.~\ref{fig:structureoptions}(c,d), with details listed in Tables~\ref{t:pmft} and \ref{t:pmmsg}. While both structures have the same magnetic space group, they are symmetry distinct, in the sense that there is not a symmetry operation of the parent cubic space group relating these structures. This can be seen most clearly from the fact that the ``+--" structure has octahedra where all spins are pointing in (and vice versa) whereas the ``++" structure does not have any such octahedra. Despite this difference, a calculation of the energy due to dipole-dipole interactions or staggered DM interactions is the same in both cases, so there is no clear physical argument for which is more likely.

\begin{longtable}{l c c c}
    \caption{The +-- magnetic structure, shown in terms of Fourier coefficients for each Tb ion in the primitive unit cell of the parent $Im\bar{3}$ structure. The corresponding magnetic R-factor is $R_{\text{mag}}=11.6\%$.\label{t:pmft}}\\
    \hline
    \hline
    \endfirsthead
    \caption{(continued)}\\
    \hline
    \hline
    \endhead
    \hline
    \hline
    \endfoot
    \hline
    \hline
    \endlastfoot
    Atom$(x,y,z)$/propagation vector & $M_x$ ($\mu_{\rm{B}}$) & $M_y$ ($\mu_{\rm{B}}$) & $M_z$ ($\mu_{\rm{B}}$) \\
    \hline
     Tb1\_1(0,0.1915,0.3029)/$\bm{k}_1$    & 4.62(9) & 0 & 0 \\
     Tb1\_2(0,-0.1915,0.3029)/$\bm{k}_1$ & 4.62(9) & 0 & 0 \\
     Tb1\_3(0,-0.1915,-0.3029)/$\bm{k}_1$ & 4.62(9) & 0 & 0 \\
     Tb1\_4(0,0.1915,-0.3029)/$\bm{k}_1$ & 4.62(9) & 0 & 0 \\
     \hline
     Tb2\_1(0.3029,0,0.1915)/$\bm{k}_1$ & 0 & -4.62(9) & 0 \\
     Tb2\_2(-0.3029,0,0.1915)/$\bm{k}_1$ & 0 & -4.62(9) & 0 \\
     Tb2\_3(-0.3029,0,-0.1915)/$\bm{k}_1$ & 0 & -4.62(9) & 0 \\
     Tb2\_4(0.3029,0,-0.1915)/$\bm{k}_1$ & 0 & -4.62(9) & 0 \\
     \hline
     Tb3\_1(0.1915,0.3029,0)/$\bm{k}_2$ & 0 & 0 & 10.2(2) \\
     Tb3\_2(-0.1915,-0.3029,0)/$\bm{k}_2$ & 0 & 0 & -10.2(2) \\
     \hline
     Tb4\_1(-0.1915,0.3029,0)/$\bm{k}_2$ & 0 & 0 & 10.2(2) \\
     Tb4\_2(0.1915,-0.3029,0)/$\bm{k}_2$ & 0 & 0 & -10.2(2)
\end{longtable}

 \begin{longtable}{l  l}
        \caption{The ``+--" structure, described in terms of its magnetic space group. The atomic positions are fixed to those from the 30~K refinement. Note that the Tb site labels are unrelated to those used in Table~\ref{t:pmft} and that the standard setting for the magnetic space group is used, so that the definitions of the axes and propagation vectors do not correspond to those given in the main text and in the representational analysis above.\label{t:pmmsg}}\\
    \hline
    \hline
    \endfirsthead
    \caption{(continued)}\\
    \hline
    \hline
    \endhead
    \hline
    \hline
    \endfoot
    \hline
    \hline
    \endlastfoot
    Parent space group & $Im\bar{3}.1^{\prime}$ \\
    Propagation vector(s) & $(100)$, $(\frac{1}{2},0,\frac{1}{2})$ \\
    \makecell[tl]{Transformation from parent basis to \\ the one used} & $a+c,b,-a+c;0,0,\frac{1}{2}$ \\
    MSG symbol (BNS) & $P_C 2_1/c$ \\
    MSG number (BNS) & 14.84 \\
    MSG symbol (UNI) & $P 2_1/c.1^{\prime}_C[C 2/c]$ \\
    \makecell[tl]{Transformation from basis used to \\standard setting of MSG} & $a,b,c;0,0,0$ \\
    Magnetic point group & $2/m.1^{\prime}$ \\
    Active irreps & $\mathrm{m}N_2^-$, primary order parameter \\
    & $\mathrm{m}H_4^+$, special direction, primary order parameter \\
    Unit cell parameters (\AA, $^{\circ}$) & \makecell[tl]{$a=21.91182, b=15.49400, c=21.91182,$ \\ $\alpha=90.0, \beta=90.0, \gamma=90.0$}\\
    MSG symmetry operations (4) & $x,y,z,+1$ \\
    & $-x, y+\frac{1}{2}, -z+\frac{1}{2},+1$ \\
    & $-x, -y, -z, +1$ \\
    & $x, -y+\frac{1}{2}, z+\frac{1}{2}, +1$\\
    MSG symmetry centering operations (2) & $x,y,z,+1$ \\
     & $x+\frac{1}{2}, y+\frac{1}{2},z,-1$ \\
     Positions and occupancies of magnetic atoms & Tb1 Tb 0.34575 0.80288 0.65425 1.0\\
     & Tb2 Tb 0.84575 0.80288 0.15425 1.0\\
     & Tb3 Tb 0.40144 0.69149 -0.09856 1.0\\
     & Tb4 Tb -0.09856 0.69149 0.40144 1.0\\
     & Tb5 Tb 0.49719 0.50000 0.69431 1.0\\
     & Tb6 Tb 0.80569 0.50000 0.00281 1.0\\
     Positions and occupancies of non-magnetic atoms & Cd1 Cd 0.47822 0.70237 0.63996 1.0\\
     & Cd2 Cd -0.02178  0.70237  0.13996 1.0\\
     & Cd3 Cd  0.63996  0.29763  0.47822 1.0\\
     & Cd4  Cd 0.13996  0.29763 -0.02178 1.0\\
    & Cd5  Cd  0.52031  0.61819  0.81795 1.0\\
    & Cd6  Cd  0.02031  0.61819  0.31795 1.0\\
    & Cd7  Cd  0.81795  0.38181  0.52031 1.0\\
    & Cd8  Cd  0.31795  0.38181  0.02031 1.0\\
    & Cd9  Cd  0.41027  0.83826  0.79209 1.0\\
    & Cd10 Cd -0.08973  0.83826  0.29209 1.0\\
    & Cd11 Cd  0.70791  0.16174  0.08973 1.0\\
    & Cd12 Cd  0.20791  0.16174  0.58973 1.0\\
    & Cd13 Cd  0.27956  0.60280  0.72044 0.333\\
    & Cd14 Cd  0.77956  0.60280  0.22044 0.333\\
    & Cd15 Cd  0.33096  0.50000  0.72816 0.333\\
    & Cd16 Cd  0.72816  0.50000  0.33096 0.333\\
    & Cd17 Cd  0.30140  0.55913  0.80140 0.333\\
    & Cd18 Cd  0.80140  0.55913  0.30140 0.333\\
    & Cd19 Cd  0.36954  0.59558  0.63046 1.0\\
    & Cd20 Cd  0.86954  0.59558  0.13046 1.0\\
    & Cd21 Cd  0.41733  0.50000  0.82175 1.0\\
    & Cd22 Cd  0.82175  0.50000  0.41733 1.0\\
    & Cd23 Cd  0.29779  0.73909  0.79779 1.0\\
    & Cd24 Cd  0.79779  0.73909  0.29779 1.0\\
    & Cd25 Cd  0.45337  0.84783  0.54663 1.0\\
    & Cd26 Cd -0.04663  0.84783  0.04663 1.0\\
    & Cd27 Cd  0.62728  0.50000  0.77946 1.0\\
    & Cd28 Cd  0.77946  0.50000  0.62728 1.0\\
    & Cd29 Cd  0.42391 -0.09325 -0.07609 1.0\\
    & Cd30 Cd -0.07609 -0.09325  0.42391 1.0\\
    & Cd31 Cd  0.40482  0.65482  0.75000 1.0\\
    & Cd32 Cd -0.09518  0.65482  0.25000 1.0\\
    & Cd33 Cd  0.75000  0.34518  0.40482 1.0\\
    & Cd34 Cd  0.25000  0.34518 -0.09518 1.0\\
    & Cd35 Cd  0.34762  0.00000  0.84762 1.0\\
    & Cd36 Cd  0.50000  0.69525  0.50000 1.0\\
    & Cd37 Cd  0.59762  0.50000 -0.09762 1.0\\
    & Cd38 Cd  0.45243  0.50000 -0.04757 1.0\\
    & Cd39 Cd  0.75000  0.09514  0.25000 1.0\\
    & Cd40 Cd  0.75000 -0.09514  0.25000 1.0\\
    & Cd41 Cd  0.45243  0.50000  0.54757 1.0\\
    \makecell[tl]{Magnetic moment components ($\mu_{\rm{B}}$) \\
    of magnetic atoms, symmetry \\ 
    constraints and moment magnitudes} &
    \makecell[tl]{
    Tb1   3.27  0.00000   3.27 Mx,My,Mz 4.62(9) \\
Tb2   3.27  0.00000   3.27 Mx,My,Mz 4.62(9) \\
Tb3  -3.27  0.00000  3.27 Mx,My,Mz 4.62(9) \\
Tb4  -3.27  0.00000  3.27 Mx,My,Mz 4.62(9) \\
Tb5   0.00000  10.2(2)  0.00000 Mx,My,Mz 10.2(2) \\
Tb6   0.00000 -10.2(2)  0.00000 Mx,My,Mz 10.2(2)}
 \end{longtable}

There is also another possibility for the structure, which is that two arms of the star of $\bm{k}_2=$ \hho  could be active in the structure, i.e., \hho on orbits Tb3 and Tb4 and $(0\frac{1}{2}\frac{1}{2})$ on orbit Tb1, with $(100)$ only having non-zero Fourier coefficients on orbit Tb2. This structure has much lower symmetry (magnetic space group $P_S\bar{1}$) with many magnetic irreps allowed to be active, for which we see no evidence. We therefore rule out this structure from further consideration. 

\subsection{Magneto-elastic coupling}
We have verified that the magnetic structure determined via representational analysis relative to $Im\bar{3}$ is compatible with the previously determined monoclinic distortion which arises from orientational ordering of the Cd tetrahedra at the center of the Tsai type clusters \cite{Tamura2005a,Nishimoto2013}, described in Sec.~\ref{sec:parent}. Specifically, this monoclinic distortion is via the $N_1^-$ irrep of $Im\bar{3}.1^{\prime}$, with the resulting space group being $C2/c.1^{\prime}$. We have performed several tests of the symmetry compatibility.

\paragraph*{Resulting space group} The magnetic space group that we find using ISOCIF \cite{Isotropy} is $P_C2_1/c$ in the BNS notation or $P2_1/c . 1^{\prime}_C [C2/c]$ in the UNI notation. The resulting non-magnetic spacegroup is therefore $C2/c$, as determined in Ref.~\cite{Nishimoto2013}.

\paragraph*{Induced order parameters} The induced order parameters were calculated using ISODISTORT \cite{Isotropy, Campbell2006}. The refined structure, in which the only active modes are m$N_2^-$ and m$H_4^+$ induces the $N_1^-$ irrep, i.e., the isotropy subgroup of the $N_1^-$ irrep is a supergroup of the isotropy subgroup in which both m$N_2^-$ and m$H_4^+$ are active. This means that the former can be a high temperature parent of the latter. It should be noted that the induced $N_1^-$ corresponds to a different arm of the star compared to m$N_2^-$, $(\frac{1}{2},\frac{1}{2},1)$ compared to $(\frac{1}{2},\frac{1}{2},0)$. In fact, any two of m$N_2^-$, m$H_4^+$ and $N_1^-$ can be taken as the primary order parameters, meaning that if two of them are active, the third will be induced. Since we know that the $N_1^-$ irrep is active at temperatures well above the magnetic phase transition, this explains why both magnetic irreps become active at the same temperature, since the presence of one induces the other in the presence of $N_1^-$.

\paragraph*{Free energy invariants} 
The symmetry relations between the structural distortion and magnetic order parameters, discussed above, have further support from the analysis of the free energy invariants, which were calculated using ISOTROPY \cite{Isotropy}. It was found that there is a trilinear term which couples all three irreps $N_1^-$, m$N_2^-$ and m$H_4^+$. This invariant remains non-zero even upon restricting to the relevant special directions:
that is, only one arm of the star contributes for each of $N_1^-$ and m$N_2^-$, $(\frac{1}{2}\frac{1}{2}1)$ and \hho respectively, and the order parameter direction for m$H_4^+$ is $(a,b,0)$.

\section{Proposed exchange interactions}

\begin{figure}[h]
    \centering
    \includegraphics[width=0.5\linewidth]{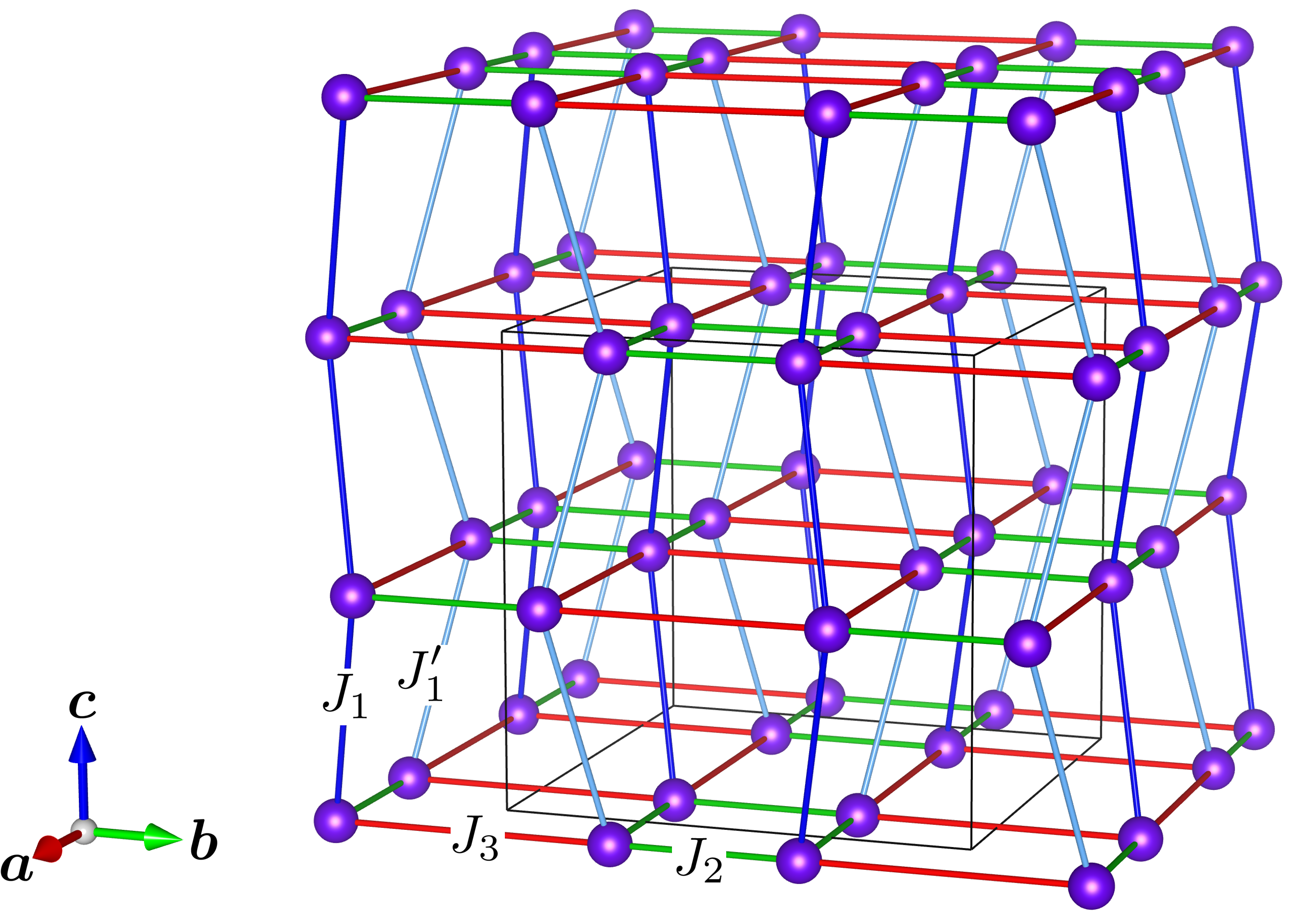}
    \caption{The full three-dimensional model of proposed Heisenberg exchange interactions for the apical spins in \ch{Cd6Tb}. The interactions for the equatorial spins are obtained by the appropriate 3-fold rotations and setting $J_1 = J_1^{\prime}$.}
    \label{fig:interactions_3d}
\end{figure}

The main text presented a two-dimensional version of the proposed exchange interactions for simplicity. In Fig.~\ref{fig:interactions_3d}, we show the full three-dimensional model. While the $J_2$ and $J_3$ bonds along $a$ in this figure are not symmetry-equivalent to those along $b$, it is possible to reproduce the observed ordering patterns while setting them equal to each other, so this is considered a suitable minimal model for the interactions.
 
\section{$\mu$SR measurements}

\subsection{Experimental details}
A 0.5~g powder sample of \ch{Cd6Tb} was enclosed between two pieces of Kapton tape. The sample was placed in a helium flow cryostat with a base temperature of 1.5~K and measurements were taken at a series of temperatures including 5~K. The temperature-dependence of the data will be reported in a future work. 

During the experiment, some helium gas entered the pocket between the pieces of Kapton tape and caused it to expand, such that the powder fell to the bottom of the pocket and so no longer covered the whole muon beam area. This meant that a sizeable fraction of the muons stopped outside the sample in the data taken after this happened. It was possible to determine that the helium gas entered the pocket between when the 5~K and the 1.5~K data were taken, because the veto rate changed significantly. 

\subsection{Analysis}

The analysis of all $\mu$SR data was carried out using MuSRfit \cite{Suter2012}. 

\subsubsection{Data at 5~K}

\begin{figure}
    \centering
    \includegraphics[width=0.8\linewidth]{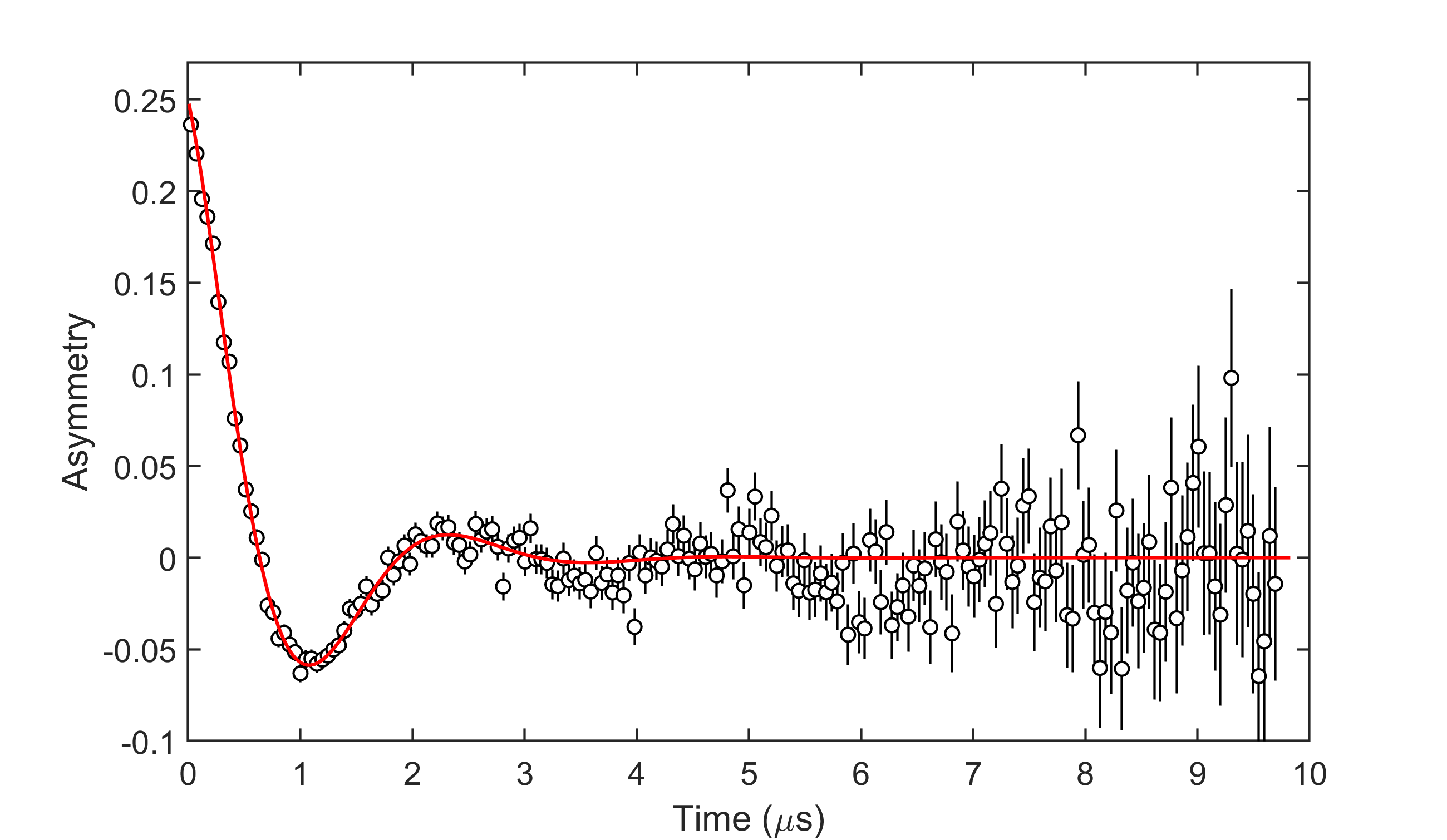}
    \caption{Calibration data taken at 300~K in a 30~G transverse field at the beginning of the $\mu$SR experiment. The solid line corresponds to Eq.~(\ref{eq:calib1}) with parameters in Table~\ref{t:mucalib1}.}
    \label{fig:musr_calib1}
\end{figure}

The data at 5~K were collected before the helium gas entered the pocket, so the calibration performed at 300~K and 30~G at the beginning of the experiment could be used. These calibration data are shown in Fig.~\ref{fig:musr_calib1}. It can be seen that even at this elevated temperature, the oscillations in the applied field are strongly damped, implying that there are strong spin fluctuations on the time scales to which muons are sensitive. The calibration data were fitted to a form
\begin{equation}\label{eq:calib1}
    A(t) = A_{\text{total}}e^{-\lambda t}\cos\left(\gamma_{\mu} B t + \phi\right),
\end{equation}
where $A(t)$ is the asymmetry as a function of time, $A_{\text{total}}$ is the total initial asymmetry, $\lambda$ parameterizes the decay in the signal, $\gamma_{\mu}$ is the gyromagnetic ratio of the the muon Larmor precession frequency to the magnetic flux density, $B$ is the applied field and $\phi$ is the initial phase of the muon spin relative to the positron detector direction.
Allowing for a small non-decaying oscillating component of the signal yielded a size for this component of at most 2\% and made little difference to the quality of the fit, so this has not been included. Because the signal is so strongly decaying, we fixed $A_{\text{total}}$ to the instrumental value of 0.25, as it could not be accurately determined by fitting. The fitted parameters are shown in Table~\ref{t:mucalib1}. The $\alpha$ parameter, characterizing the sample geometry and the efficiency of the detectors, determined from this fit was used to analyze the data at 5~K.

\begin{longtable}{l l}
\caption{Parameters used to fit the initial calibration data taken at 300~K in 30~G transverse field, as used in Eq.~(\ref{eq:calib1}). \label{t:mucalib1}}\\
    \hline
    \hline
    \endfirsthead
    \caption{(continued)}\\
    \hline
    \hline
    \endhead
    \hline
    \hline
    \endfoot
    \hline
    \hline
    \endlastfoot
    $\alpha$ & 0.9932(17) \\
    $A_{\text{total}}$ & 0.25 \\
    $\lambda$ & 1.24(2) $\mu$s$^{-1}$\\
    $B$ & 29.7(5) G \\
    $\phi$ & -2.5(1.7) $^{\circ}$
\end{longtable}

To fit the data in Fig.~4 of the main text, the following form was used:
\begin{equation}\label{eq:mu5K}
    A(t) = A_{\text{total}}\left[f_{\text{magnetic}}\left(\frac{2}{3}e^{-\lambda_T}\cos{\left(\gamma_{\mu}B_{\text{int}}t\right)}  +\frac{1}{3}\left[f_1e^{-\lambda_1t} + \left(1-f_1\right)e^{-\lambda_2t} \right]\right) + \left(1-f_{\text{magnetic}}\right)\right].
\end{equation}
where $A_{\text{total}}$ was again fixed to 0.25, and $\alpha$ was fixed to the value listed in Table~\ref{t:mu300K}.
This form allows for a constant contribution controlled by $(1-f_{\text{magnetic}})$, assumed to be due to muons landing in the sample holder or in regions of the sample where the magnetic fields are negligible. The oscillating fraction was fixed to 2/3 as, in a powder sample, this is the average proportion of the internal field which is perpendicular to the initial muon direction and can therefore cause precession. The best fit parameters for the 5~K data are given in Table~\ref{t:mu5K}.

The decay in the oscillating component could be due either to dynamical effects or to there being a range of different static fields present at different muon stopping sites. The decay in the longitudinal component, however, can be attributed to dynamics.

\begin{longtable}{l l}
\caption{Parameters in Eq.~(\ref{eq:mu5K}) corresponding to the data at 5~K shown in Fig.~4 of the main text. Parameters with no listed uncertainty were fixed to the given values.\label{t:mu5K}}.\\
    \hline
    \hline
    \endfirsthead
    \caption{(continued)}\\
    \hline
    \hline
    \endhead
    \hline
    \hline
    \endfoot
    \hline
    \hline
    \endlastfoot
    $\alpha$ & 0.9932 \\
    $A_{\text{total}}$ & 0.25 \\
    $f_{\text{magnetic}}$ & 0.957(7) \\
    $f_1$ & 0.59(3) \\
    $\lambda_T$ & 161(11) $\mu$s$^{-1}$\\
    $\lambda_1$ & 0.059(17) $\mu$s$^{-1}$ \\
    $\lambda_2$ & 1.3(2) $\mu$s$^{-1}$ \\
    $B$ & $1.22(4)\times 10^4$ G \\
    $\phi$ & 35(12) $^{\circ}$
\end{longtable}

\subsubsection{Data at 1.5~K}

\begin{figure}[h]
    \centering
    \includegraphics[width=0.8\linewidth]{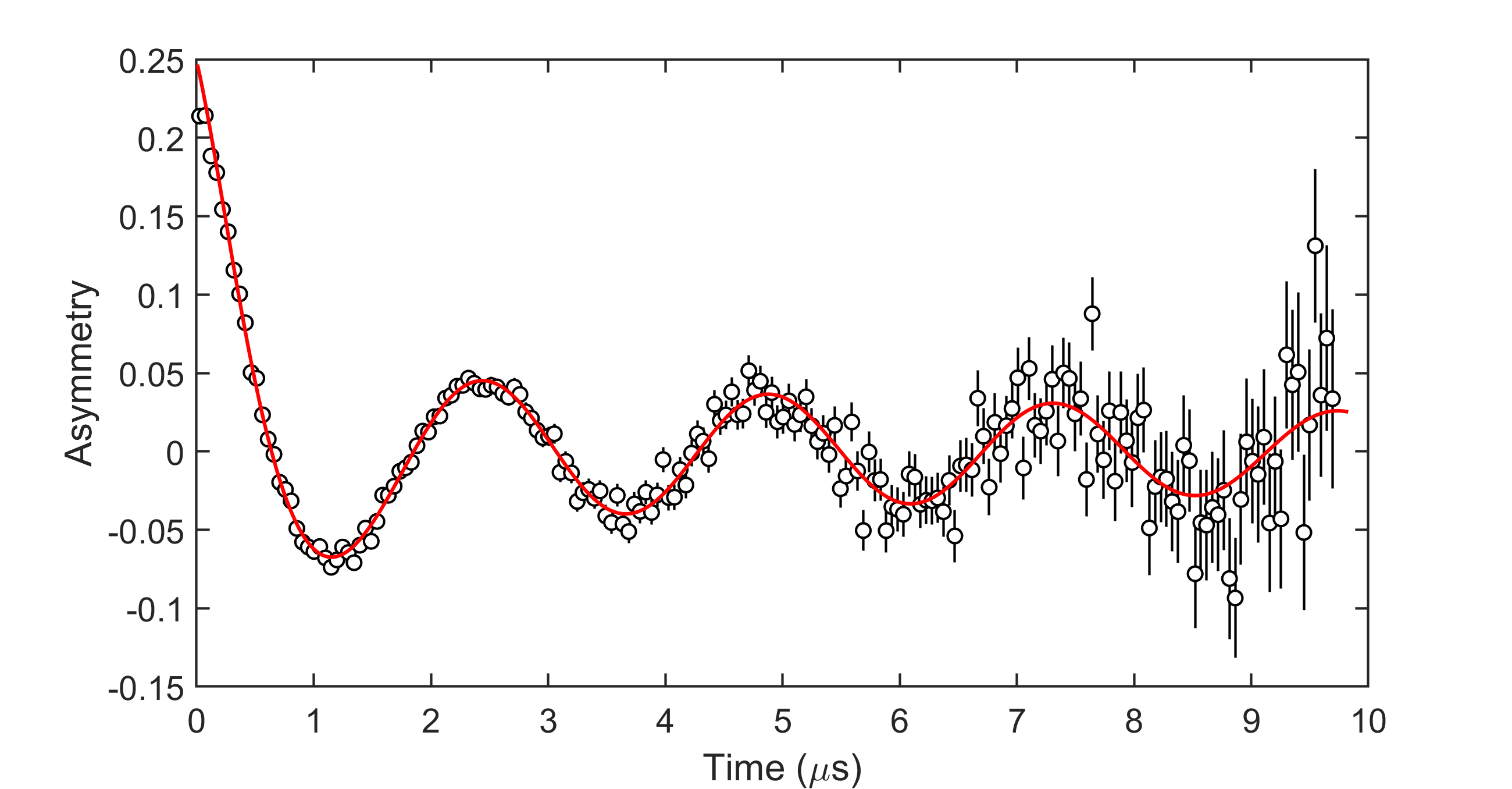}
    \caption{$\mu$SR data taken at 300~K with a 30~G transverse field. The solid red curve is given by Eq.~(\ref{eq:mu300K}) with the parameters in Table~\ref{t:mu300K}.}
    \label{fig:musr_calib2}
\end{figure}

Because of the gas entering the sample pocket, a second calibration measurement at 300~K in 30~G transverse field was taken at the end of the experiment. This was used to determine the appropriate $\alpha$ as the geometry had changed, as well as to obtain an estimate of the proportion of muons which stopped outside the sample, $(1-f_{\text{magnetic}})$. 
These data were fitted to a form 
\begin{equation}\label{eq:mu300K}
    A(t) = A_{\text{total}}\left[f_{\text{magnetic}}e^{-\lambda_1t}\cos{\left(\gamma_{\mu}Bt+\phi\right)}+\left(1-f_{\text{magnetic}}\right)e^{-\lambda_2t}\cos{\left(\gamma_{\mu}Bt+\phi\right)}\right],
\end{equation}
where the total initial asymmetry $A_{\text{total}}$ was again fixed to 0.25. The resulting fit is shown in Fig.~\ref{fig:musr_calib2} and the fitted parameters are given in Table~\ref{t:mu300K}.

\begin{longtable}{l l}
\caption{Parameters in Eq.~(\ref{eq:mu300K}). \label{t:mu300K}}\\
    \hline
    \hline
    \endfirsthead
    \caption{(continued)}\\
    \hline
    \hline
    \endhead
    \hline
    \hline
    \endfoot
    \hline
    \hline
    \endlastfoot
    $\alpha$ & 1.0868(13) \\
    $A_{\text{total}}$ & 0.25 \\
    $f_{\text{magnetic}}$ & 0.724(17) \\
    $\lambda_1$ & 2.28(11) $\mu$s$^{-1}$\\
    $\lambda_2$ & 0.15(2) $\mu$s$^{-1}$ \\
    $B$ & 30.56(13) G \\
    $\phi$ & -6(1) $^{\circ}$
\end{longtable}

Due to the difference between the calibration signals at the beginning and end of the experiment (Figs.~\ref{fig:musr_calib1} and \ref{fig:musr_calib2}, respectively), it was assumed that the fraction with the fast decay corresponded to the fraction landing in the sample. 

To obtain alternative estimates of $f_{\text{magnetic}}$ in the regime after the gas entered the pocket, data at temperatures above the ordering temperature were inspected, where no static component of the signal is expected. This yielded the value of $f_{\text{magnetic}}$ given above as the minimum estimate, and a maximum estimate of 0.842(2). 

\begin{figure}[h]
    \centering
    \includegraphics[width=\linewidth]{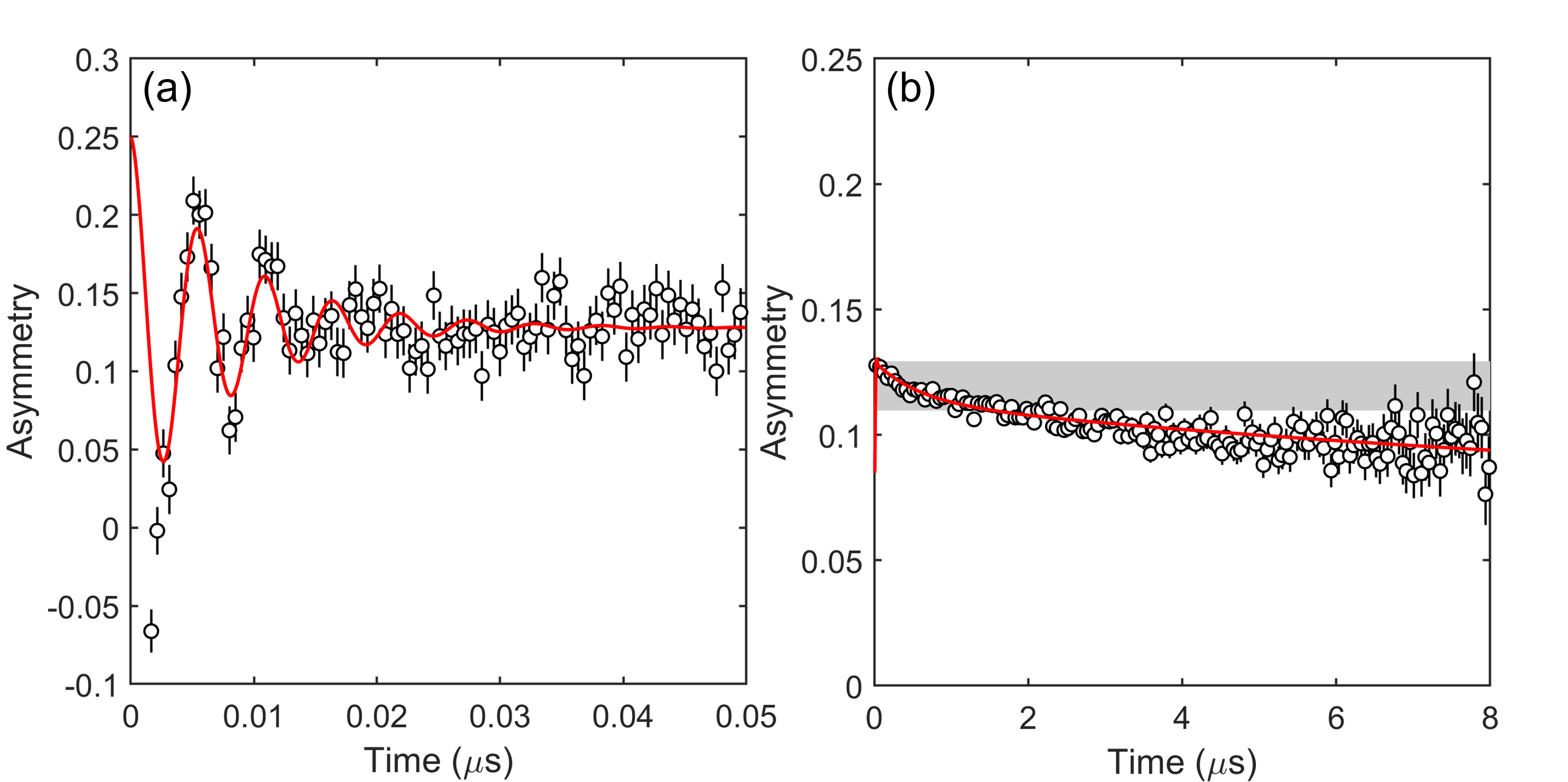}
    \caption{$\mu$SR data at 1.5~K in zero field showing both the short time oscillatory behaviour (a) and the long time decay (b). The gray shaded region in (b) indicates the range of possible values for $A_{1/3}$ obtained via different estimates for $f_{\text{magnetic}}$}
    \label{fig:musr_1K5}
\end{figure}

Data taken in zero field at 1.5~K are shown in Fig.~\ref{fig:musr_1K5}. The behavior qualitatively resembles what is seen at 5~K.
These data were therefore fitted to the same form as the 5~K data, Eq.~(\ref{eq:mu5K})
where $A_{\text{total}}$ was again fixed to 0.25, and $\alpha$ and $f_{\text{magnetic}}$ were fixed to the values listed in Table~\ref{t:mu300K}. The fitted parameters are given in Table~\ref{t:mu1.5K}. 

\begin{longtable}{l l}
\caption{Parameters in Eq.~(\ref{eq:mu5K}) used to fit the data at 1.5~K. Parameters with no listed uncertainty were fixed to the given values.\label{t:mu1.5K}}.\\
    \hline
    \hline
    \endfirsthead
    \caption{(continued)}\\
    \hline
    \hline
    \endhead
    \hline
    \hline
    \endfoot
    \hline
    \hline
    \endlastfoot
    $\alpha$ & 1.0868 \\
    $A_{\text{total}}$ & 0.25 \\
    $f_{\text{magnetic}}$ & 0.724 \\
    $f_1$ & 0.266 \\
    $\lambda_T$ & 122(10) $\mu$s$^{-1}$\\
    $\lambda_1$ & 1.7(4) $\mu$s$^{-1}$ \\
    $\lambda_2$ & 0.072(13) $\mu$s$^{-1}$ \\
    $B$ & $1.346(15)\times 10^4$ G \\
    $\phi$ & 0 $^{\circ}$
\end{longtable}

To determine whether these data indicate persistent dynamics or whether they could be consistent with static disorder, the position of the 1/3 tail was determined according to
\begin{equation}
    A_{1/3} = A_{\text{total}}\left(1-\frac{2}{3}f_{\text{magnetic}}\right).
\end{equation}
The range of values of $A_{1/3}$ corresponding to different estimates of $f_{\text{magnetic}}$ are shown as the gray shaded area in Fig.~\ref{fig:musr_1K5}(b). The data clearly decay past the lowest estimate for $A_{1/3}$, strongly suggesting that the magnetism at 1.5~K, like at 5~K, is dynamic on time scales probed by $\mu$SR.

We note that there is a magnetic transition at 2.4~K, as previously evidenced in heat capacity and magnetic susceptibility \cite{Tamura2010}. This transition could therefore perhaps be supposed to be a spin freezing transition from a dynamically disordered state at 5~K to a statically disordered one, and indeed, the long time behaviour of the $\mu$SR signal is certainly flatter at 1.5~K than at 5~K. This indicates slower fluctuation dynamics as the relaxation rate of the tail is directly proportional to the fluctuation rate in the slow fluctuation limit. In the following section however, we present AC susceptibility data which show no evidence for a spin freezing transition, providing further evidence that the disorder remains dynamic down to 1.5~K.

\section{AC susceptibility}
\subsection{Experimental details}

AC susceptibility measurements were performed using a Quantum Design Magnetic Property Measurement System 3 on a 41.9 mg powder sample of \ch{Cd6Tb}. Measurements were performed at several frequencies.

\subsection{Results}

In the case of a spin-glass transition, there are two main signatures which would be expected to be seen in AC susceptibility: (i) at temperatures below the transition, the imaginary component of the susceptibility would become non-zero and (ii) the temperature at which the cusp in the susceptibility occurs would increase with increasing frequency.

\begin{figure}
    \centering
    \includegraphics[width=\linewidth]{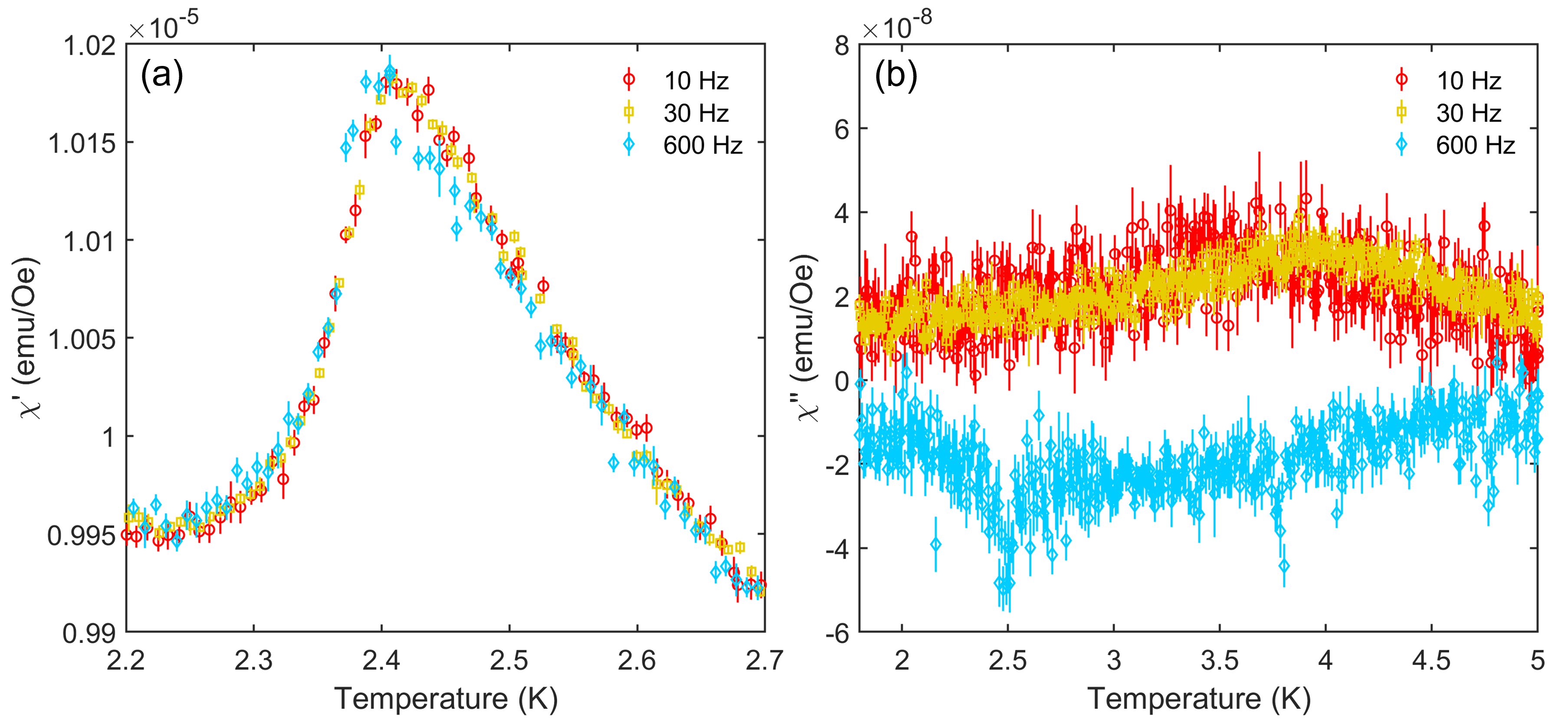}
    \caption{(a) The real part and (b) the imaginary part of the AC susceptibility of \ch{Cd6Tb} measured at various frequencies.}
    \label{fig:ac}
\end{figure}

The results of the AC susceptibility measurement are shown in Fig.~\ref{fig:ac}. The measurement of the cusp in the real part of the susceptibility [Fig.~\ref{fig:ac}(a)] shows that if anything, the cusp moves to \emph{lower} temperatures as the frequency increases, which is the opposite of what is expected at a spin freezing transition. Meanwhile, the imaginary part of the susceptibility [Fig.~\ref{fig:ac}(b)] is negligible both above and below the transition. These measurements therefore indicate that the transition at 2.4~K is \emph{not} a spin freezing transition. Taken together with the facts that the $\mu$SR data indicate that the disorder at 5~K is dynamic, and the neutron diffraction data indicate that disorder persists at 1.5~K, this provides strong evidence that the disorder at 1.5~K is also dynamic.